\documentclass[vecphys]{svmult}

\def\be{\begin{equation}}
\def\ee{\end{equation}}
\def\bea{\begin{eqnarray}}
\def\eea{\end{eqnarray}}

\usepackage{makeidx}         
\usepackage{graphicx}        
\usepackage{multicol}        
\usepackage[bottom]{footmisc}

\makeindex             

\begin{document}

\title*{Conceptual Problems of Inflationary Cosmology and a New Approach to
Cosmological Structure Formation}

\titlerunning{Problems of Inflation} 

\author{Robert H. Brandenberger}

\institute{Department of Physics, McGill University, Montreal, QC, H3A 2T8,
Canada
\texttt{rhb@hep.physics.mcgill.ca}}
%
%
\maketitle

\begin{abstract}

In spite of its great phenomenological success, current models
of scalar field-driven inflation suffer from important unresolved
conceptual issues. New fundamental physics will be required to 
address these questions. String theory is a candidate for a unified
quantum theory of all four forces of nature. As will be shown,
string theory may lead to a cosmological background quite different
from an inflationary cosmology, and may admit a new stringy mechanism
for the origin of a roughly scale-invariant spectrum of 
cosmological fluctuations.

\end{abstract}

\section{Introduction}
\label{sec:1}

The inflationary universe scenario \cite{Guth} (see also 
\cite{Sato,Starob1,Brout} for earlier ideas) has been extremely successful
phenomenologically. In addition to providing answers to some
key open questions of Standard Big Bang cosmology such as
the horizon, flatness and entropy problems, inflation gives rise
to a causal mechanism for structure formation \cite{Mukh} (see
also \cite{Press} for more qualitative arguments and
\cite{Starob2} for an early computation of the spectrum of gravitational
waves in an inflationary background). Quantum vacuum fluctuations
during the period of exponential expansion lead to a roughly
scale-invariant spectrum of (in the simplest models) adiabatic
fluctuations. These fluctuations are squeezed while their wavelength
is larger than the Hubble radius, and thus re-enter the Hubble radius
at late times as standing waves. As realized a long time ago in
\cite{Sunyaev,Peebles}, these features predict ``acoustical"
oscillations in the angular power spectrum of cosmic microwave
background anisotropies. Both the approximate scale-invariance
and the acoustical oscillations of the spectrum have recently,
many years after these features were predicted, been confirmed
by CMB anisotropy experiments \cite{COBE,Boomerang,WMAP}.

On the theoretical front, the situation is much less satisfactory.
In spite of over twenty years of research, no convincing theory of
inflation has emerged. There are many models of inflation, but all
of them involve new scalar fields. String theory and most other theories
beyond the standard model do predict scalar fields, and thus may
well eventually give rise to a good theory of inflation (see
e.g. \cite{Burgess,Cline,Linde} for reviews on avenues to obtain
inflation in string theory), but at the moment the question is
not resolved. Furthermore, some of the conceptual
issues which will be raised below (Section 2) are generic to 
any implementation of inflation by scalar fields in the context of 
Einstein gravity.

Thus, it is important to keep an open mind to the possibility that
an early universe scenario which does not involve a period of
cosmological inflation will emerge. As will be shown below,
the new degrees of freedom and new symmetries of string theory
give rise to the possibility of a cosmological background very
different from that of inflationary cosmology (Section 3). Within this 
background cosmology, a stringy mechanism which can generate
a scale-invariant spectrum of cosmological perturbations has
recently been proposed \cite{NBV} (see also \cite{Ali,BNPV2} for
reviews). This mechanism, which yields a distinctive signature,
namely a slight blue tilt in the spectrum of gravitational waves
\cite{BNPV1}, will be discussed in Section 4. Subsection 4.5
reviews some results which were completed after the Colloque
in Paris and appeared in \cite{Betal}.


\section{Problems of Scalar Field-Driven Inflation}

\subsection{Review of the Inflationary Universe Scenario}
 
Before discussing some key conceptual problems of 
conventional scalar field-driven
inflationary cosmology, let us recall some of the main features of
cosmological inflation. To set our notation, we use the following
metric for the homogeneous and isotropic background space-time:
\begin{equation}
ds^2 \, = \, dt^2 - a(t)^2 d{\bf x}^2 \, ,
\end{equation}
where $t$ is physical time, ${\bf x}$ denote comoving coordinates
on the spatial sections which we for simplicity assume to be ${\cal R}^3$, 
and $a(t)$ is the scale factor.

Figure 1 is a sketch of the space-time
structure of an inflationary universe. The vertical axis is time,
the horizontal axis is physical length. The time period between
$t_i$ and $t_R$ is the period of inflation (here for simplicity taken
to be exponential). During the period of inflation, the Hubble radius
\begin{equation}
l_H(t) \, \equiv \, H^{-1}(t) \,\,\, {\rm where} \,\,\,
H(t) \, \equiv \, {\dot a(t)} / a(t)
\end{equation}
is constant. After inflation,
the Hubble radius increases linearly in time. In contrast, the
physical length corresponding to a fixed co-moving scale increases
exponentially during the period of inflation, and then grows either
as $t^{1/2}$ (radiation-dominated phase) or $t^{2/3}$ (matter-dominated
phase), i.e. less fast than the Hubble radius. 
\begin{figure}
\centering
\includegraphics[height=8cm]{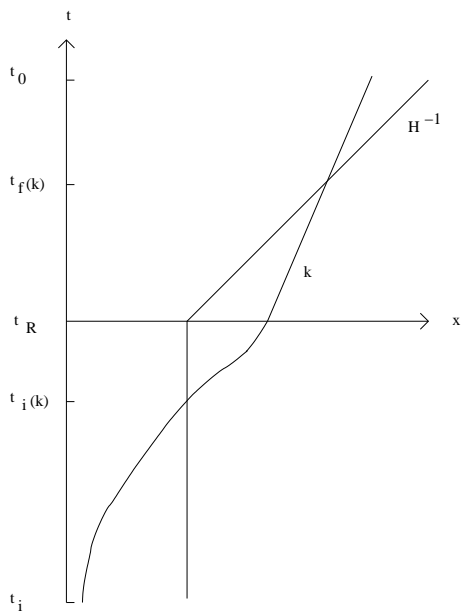}
\caption{Space-time diagram (sketch) showing the evolution
of scales in inflationary cosmology. The vertical axis is
time, and the period of inflation lasts between $t_i$ and
$t_R$, and is followed by the radiation-dominated phase
of standard big bang cosmology. During exponential inflation,
the Hubble radius $H^{-1}$ is constant in physical spatial coordinates
(the horizontal axis), whereas it increases linearly in time
after $t_R$. The physical length corresponding to a fixed
comoving length scale labelled by its wavenumber $k$ increases
exponentially during inflation but increases less fast than
the Hubble radius (namely as $t^{1/2}$), after inflation.}
\label{fig:1}       
\end{figure}

The key feature of inflationary cosmology which can be seen from
Figure 1 is the fact that fixed comoving scales are red-shifted
exponentially relative to the Hubble radius during the period of
inflation. Provided that the period of inflation lasted more than
about $50$ Hubble expansion times (this number is obtained assuming
that the energy scale of inflation is of the order of $10^{16}$GeV),
then modes with a wavelength today comparable to the current Hubble radius 
started out at the beginning of the period of inflation with a wavelength 
smaller than the Hubble radius at that time. Thus, it is possible to imagine
a microscopic mechanism for creating the density fluctuations in the
early universe which evolve into the cosmological structures we observe
today.

Since during the period of inflation any pre-existing ordinary matter
fluctuations are red-shifted, it is reasonable to assume that quantum
vacuum fluctuations are the source of the currently observed structures
\cite{Mukh} (see also \cite{Press}). The time-translational
symmetry of the inflationary phase leads, independent of a precise
understanding of the generation mechanism for the fluctuations, to
the prediction that the spectrum of cosmological perturbations should
be approximately scale-invariant \cite{Press,Sato}. 

The quantum theory of linearized cosmological perturbations 
\cite{Sasaki,Mukh2}, in particular applied to inflationary
cosmology, has in the mean time become a well-developed research area (see 
e.g. \cite{MFB} for a detailed review, and \cite{RHBrev2} for a
pedagogical introduction). For simple scalar field matter, there is
a single canonically normalized variable, often denoted by $v$, 
which carries the information
about the ``scalar metric fluctuations'', the part of the metric
perturbations which couples at linearized level to the matter.
The equation of motion for each Fourier mode of this variable $v$ has the 
form of a harmonic oscillator with a time-dependent square mass
$m^2$, whose form is set by the cosmological background. On
scales smaller than the Hubble radius, the modes oscillate (quantum
vacuum oscillations). However, on
length scales larger than the Hubble radius, $m^2$ is negative, the
oscillations cease, and
the wave functions of these modes undergo squeezing. Since the squeezing
angle in phase space does not depend on the wave number, all modes
re-enter the Hubble radius at late times with the same squeezing angle.
This then leads to the prediction of ``acoustic'' oscillations 
\cite{Sunyaev,Peebles} in the
angular power spectrum of CMB anisotropies (see e.g.
\cite{acoustic} for a recent analytical treatment), a prediction
spectacularly confirmed by the WMAP data \cite{WMAP}, and allowing
cosmologists to fit for several important cosmological parameters.

\subsection{Nature of the Inflaton}
\label{subsec:2.1}

In the context of General Relativity as the theory of space-time,
matter with an equation of state
\be \label{eos}
p \, \simeq \, - \rho
\ee
(with $\rho$ denoting the energy density and $p$ the pressure) is
required in order to obtain almost exponential expansion of space.
If we describe matter in terms of fields with canonical kinetic
terms, a scalar field is required since in the context of usual
field theories it is only for scalar fields that a potential energy
function in the Lagrangian is allowed, and of all energy
terms only the potential energy can yield the equation of state 
(\ref{eos}). 

In order for scalar fields to generate a period of cosmological
inflation, the potential energy needs to dominate over the kinetic
and spatial gradient energies. It is generally assumed that spatial
gradient terms can be neglected. This is, however, not true in general.
Assuming a homogeneous field configuration, we must ensure that the
potential energy dominates over the kinetic energy. This leads to
the first ``slow-roll" condition. Requiring the period of inflation
to last sufficiently long leads to a second slow-roll condition, namely
that the $\ddot \varphi$ term in the Klein-Gordon equation for the
inflaton $\varphi$ be negligible. Scalar fields charged with respect
to the Standard Model symmetry groups do not satisfy the slow-roll
conditions. 

Assuming that both slow-roll conditions hold, one
obtains a ``slow-roll trajectory" in the phase space of
homogeneous $\varphi$ configurations. In large-field inflation models
such as ``chaotic inflation" \cite{chaotic}
and ``hybrid inflation" \cite{hybrid}, the slow-roll
trajectory is a local attractor in initial condition space \cite{kung}
(even when linearized metric perturbations are taken into account
\cite{hume}), whereas this is not the case \cite{goldwirth}
in small-field models such as ``new inflation" \cite{new}. As
shown in \cite{ghazalscott}, this leads to problems for some
models of inflation which have recently been proposed in the context
of string theory. To address this problem, it has been proposed
that inflation may be future-eternal \cite{eternal} and that it is
hence sufficient that there be configurations in initial condition
space which give rise to inflation within one Hubble patch, inflation
being then self-sustaining into the future. However, one must still
ensure that slow-roll inflation can locally be satisfied.

Many models of particle physics beyond the Standard Model contain a
plethora of new scalar fields. One of the most conservative extensions
of the Standard Model is the MSSM, the ``Minimal Supersymmetric Standard
Model". According to a recent study, among the many scalar fields in this
model, only a hand-full can be candidates for a slow-roll inflaton,
and even then very special initial conditions are required \cite{MSSM}.
The situation in supergravity and superstring-inspired field theories
may be more optimistic, but the issues are not settled.

\subsection{Hierarchy Problem}
\label{subsec:2.2}
 
Assuming for the sake of argument that a successful model of
slow-roll inflation has been found, one must still build in
a hierarchy into the field theory model in order to obtain an
acceptable amplitude of the density fluctuations (this is
sometimes also called the ``amplitude problem"). Unless
this hierarchy is observed, the density fluctuations will
be too large and the model is observationally ruled out.
 
In a wide class of inflationary
models, obtaining the correct amplitude requires the introduction
of a hierarchy in scales, namely \cite{Adams}
\begin{equation}
{{V(\varphi)} \over {\Delta \varphi^4}}
\, \leq \, 10^{-12} \, ,
\end{equation}
where $\Delta \varphi$ is the change in the inflaton field during
one Hubble expansion time (during inflation), and $V(\varphi)$ is
the potential energy during inflation.

This problem should be contrasted with the success of topological
defect models (see e.g. \cite{VilShell,HK,RHBrev0} for
reviews) in predicting the right oder of magnitude of
density fluctuations without introducing a new scale of physics.
The GUT scale as the scale of the symmetry breaking phase transition 
(which produces the defects) yields the correct magnitude of the
spectrum of density fluctuations \cite{CSpapers}. Topological
defects, however, cannot be the prime mechanism for the origin of
fluctuations since they do not give rise to acoustic oscillations
in the angular power spectrum of the CMB anisotropies \cite{CSanis}.

At first sight, it also does not appear to be necessary to introduce
a new scale of physics into the string gas cosmology structure formation
scenario which will be described in Section 4.
  
\subsection{Trans-Planckian Problem}

A more serious problem is the ``trans-Planckian problem" \cite{RHBrev1}.
Returning to the space-time diagram of Figure 1, we can immediately
deduce that, provided that the period of inflation lasted sufficiently
long (for GUT scale inflation the number is about 70 e-foldings),
then all scales inside of the Hubble radius today started out with a
physical wavelength smaller than the Planck scale at the beginning of
inflation. Now, the theory of cosmological perturbations is based
on Einstein's theory of General Relativity coupled to a simple
semi-classical description of matter. It is clear that these
building blocks of the theory are inapplicable on scales comparable
and smaller than the Planck scale. Thus, the key
successful prediction of inflation (the theory of the origin of
fluctuations) is based on suspect calculations since 
new physics {\it must} enter
into a correct computation of the spectrum of cosmological perturbations.
The key question is as to whether the predictions obtained using
the current theory are sensitive to the specifics of the unknown
theory which takes over on small scales.
 
One approach to study the sensitivity of the usual predictions of
inflationary cosmology to the unknown physics on trans-Planckian scales
is to study toy models of ultraviolet physics which allow explicit
calculations. The first approach which was used \cite{Jerome1,Niemeyer}
is to replace the usual linear dispersion relation for the Fourier
modes of the fluctuations by a modified dispersion relation, a
dispersion relation which is linear for physical wavenumbers smaller
than the scale of new physics, but deviates on larger scales. Such
dispersion relations were used previously to test the sensitivity
of black hole radiation on the unknown physics of the UV 
\cite{Unruh,CJ}. It was found \cite{Jerome1} that if the evolution of modes on
the trans-Planckian scales is non-adiabatic, then substantial
deviations of the spectrum of fluctuations from the usual results
are possible. Non-adiabatic evolution turns an initial state
minimizing the energy density into a state which is excited once
the wavelength becomes larger than the cutoff scale. Back-reaction
effects of these excitations may limit the magnitude of the
trans-Planckian effects, but - based on our recent study \cite{Jerome3} -
not to the extent initially assumed \cite{Tanaka,Starob3}.
Other approaches to study the trans-Planckian problem have been 
pursued, e.g. based on implementing the space-space \cite{Easther}
or space-time \cite{Ho} uncertainty relations, on a minimal length
hypothesis \cite{Kempf}, on ``minimal trans-Planckian'' assumptions (taking
as initial conditions some vacuum state at the mode-dependent time
when the wavelength of the mode is equal to the Planck scale
\cite{minimal}, or on effective field theory \cite{Cliff}, all showing
the possibility of trans-Planckian corrections (see also 
\cite{Jerome2} for a review of some of the previous work on the
trans-Planckian problem).

From the point of view of fundamental physics, the {\it trans-Planckian
problem} is not a problem. Rather, it yields a window of opportunity to
probe new fundamental physics in current and future observations, even
if the scale of the new fundamental physics is close to the Planck scale.
The point is that if the universe in fact underwent a period of inflation,
then trans-Planckian physics leaves an imprint on the spectrum of fluctuations.
The exponential expansion of space amplifies the wavelength of the 
perturbations
to observable scales. At the present time, it is our ignorance about
quantum gravity which prevents us from making any specific predictions. For
example, we do not understand string theory in time-dependent backgrounds 
sufficiently well to be able to at this time make any predictions for
observations. 

\subsection{Singularity Problem}

The next problem is the ``singularity problem". This problem, one
of the key problems of Standard Cosmology, has not been resolved in
models of scalar field-driven inflation.

As follows from the Penrose-Hawking singularity theorems of General
Relativity (see e.g. \cite{HE} for a textbook discussion), an initial
cosmological singularity is unavoidable if space-time is described in
terms of General Relativity, and if the matter
sources obey the weak energy conditions . Recently, the singularity theorems 
have been
generalized to apply to Einstein gravity coupled to scalar field
matter, i.e. to scalar field-driven inflationary cosmology \cite{Borde}.
It is shown that in this context, a past singularity at some point
in space is unavoidable. 

In the same way that the appearance of an initial singularity in Standard
Cosmology told us that Standard Cosmology cannot be the correct description of
the very early universe, the appearance of an initial singularity in
current models of inflation tell us that inflationary cosmology cannot yield
the correct description of the very, very early universe. At sufficiently
high densities, a new description will take over. In the same way
that inflationary cosmology contains late-time standard cosmology, it is
possible that the new cosmology will contain, at later times, inflationary
cosmology. However, one should keep an open mind to the possibility that
the new cosmology will connect to present observations via a route which
does not contain inflation. 

\subsection{Breakdown of Validity of Einstein Gravity}

The Achilles heel of scalar field-driven inflationary cosmology is,
however, the use of intuition from Einstein gravity at energy scales
not far removed from the Planck and string scales, scales where
correction terms to the Einstein-Hilbert term in the gravitational
action dominate and where intuition based on applying the Einstein
equations break down (see also \cite{swamp} for arguments along
these lines). 

All approaches to quantum gravity predict correction terms in the
action which dominate at energies close to the Planck scale - in
some cases in fact even much lower. Semiclassical gravity leads to
higher curvature terms, and may (see e.g. \cite{BMS,Biswas}) lead
to bouncing cosmologies without a singularity). Loop quantum
cosmology leads to similar modifications of early universe cosmology
(see e.g. \cite{Bojowald} for a recent review). String theory,
the theory we will focus on in the following sections, has a
maximal temperature for a string gas in thermal equilibrium \cite{Hagedorn},
which may lead to an almost static phase in the early universe - the
Hagedorn phase \cite{BV}. 

Common to all of these approaches to quantum gravity corrections to
early universe cosmology is the fact that a transition from a contracting (or
quasi-static) early universe phase to the rapidly expanding radiation phase
of standard cosmology can occur {\it without} violating the usual energy
conditions for matter. In particular, it is possible (as is predicted
by the string gas cosmology model discussed below) that the universe in an
early high temperature phase is almost static. This may be a common feature
to a large class of models which resolve the cosmological singularity.

Closely related to the above is the ``cosmological constant problem" for
inflationary cosmology. We know from
observations that the large quantum vacuum energy of field theories
does not gravitate today. However, to obtain a period of inflation
one is using precisely the part of the energy-momentum tensor of the 
inflaton field
which looks like the vacuum energy. In the absence of a convincing
solution of the cosmological constant problem it is unclear whether
scalar field-driven inflation is robust, i.e. whether the
mechanism which renders the quantum vacuum energy gravitationally
inert today will not also prevent the vacuum energy from
gravitating during the period of slow-rolling of the inflaton 
field \footnote{Note that the
approach to addressing the cosmological constant problem making use
of the gravitational back-reaction of long range fluctuations
(see \cite{RHBrev4} for a summary of this approach) does not prevent
a long period of inflation in the early universe.}.

\section{String Gas Cosmology}

\subsection{Preliminaries}

Since string theory is the best candidate we have
for a unified theory of all forces at the highest energies, we
will in the following explore the possible implications of
string theory for early universe cosmology.
 
An immediate problem which arises when trying to connect string theory
with cosmology is the {\it dimensionality problem}. Superstring theory
is perturbatively consistent only in ten space-time dimensions, but we
only see three large spatial dimensions. The original approach to 
addressing this problem was to assume that the six extra dimensions are
compactified on a very small space which cannot be probed with our
available energies. However, from the point of view of cosmology,
it is quite unsatisfactory not to be able to understand why it is
precisely three dimensions which are not compactified and why the compact
dimensions are stable. Brane world cosmology \cite{branereview} provides
another approach to this problem: it assumes that we live on a
three-dimensional brane embedded in a large nine-dimensional space.
Once again, a cosmologically satisfactory theory should explain
why it is likely that we will end up exactly on a three-dimensional
brane (for some interesting work addressing this issue see
\cite{Mahbub,Mairi,Lisa}).

Finding a natural solution to the dimensionality problem is thus one
of the key challenges for superstring cosmology. This challenge has
various aspects. First, there must be a mechanism which singles out
three dimensions as the number of spatial dimensions we live in.
Second, the moduli fields which describe the volume and the shape of
the unobserved dimensions must be stabilized (any strong time-dependence
of these fields would lead to serious phenomenological constraints).
This is the {\it moduli problem} for superstring cosmology. As
mentioned above, solving the {\it singularity problem} is another of
the main challenges. These are the three problems which {\it string
gas cosmology} \cite{BV,TV,ABE} explicitly addresses at the present
level of development.

In order to make successful connection with late time cosmology,
any approach to string cosmology must also solve the 
{\it flatness problem}, namely make sure that the three large
spatial dimensions obtain a sufficiently high entropy (size) to
explain the current universe. Finally, it must provide a
mechanism to produce a nearly scale-invariant spectrum of
nearly adiabatic cosmological perturbations. If string theory
leads to a successful model of inflation, then these two issues
are automatically addressed. In Section 5, we will discuss
a cosmological scenario which does not involve inflation but
nevertheless leads to a viable structure formation scenario
\cite{NBV}.

\subsection{Heuristics of String Gas Cosmology}

In the absence of a non-perturbative formulation of string theory,
the approach to string cosmology which we have suggested,
{\it string gas cosmology} \cite{BV,TV,ABE} (see also \cite{Perlt}
for early work, and \cite{BattWat,RHBrev5,RHBrev6} for reviews),
is to focus on symmetries and degrees of freedom which are new to
string theory (compared to point particle theories) and which will
be part of a non-perturbative string theory, and to use
them to develop a new cosmology. The symmetry we make use of is
{\it T-duality}, and the new degrees of freedom are 
{\it string winding modes} and {\it string oscillatory modes}.

We take all spatial directions to be toroidal, with
$R$ denoting the radius of the torus. Strings have three types
of states: {\it momentum modes} which represent the center
of mass motion of the string, {\it oscillatory modes} which
represent the fluctuations of the strings, and {\it winding
modes} counting the number of times a string wraps the torus.
Both oscillatory and winding states are special to strings.
Point particle theories do not contain these modes. 

The energy of an oscillatory mode is independent of $R$, momentum
mode energies are quantized in units of $1/R$, i.e.
\be
E_n \, = \, n {1 \over R} \, ,
\ee
whereas the winding mode energies are quantized in units of $R$, i.e.
\be
E_m \, = \, m R \, ,
\ee
where both $n$ and $m$ are integers. The energy of oscillatory modes
does not depend on $R$. 

The T-duality symmetry is the invariance of the spectrum of string
states under the change
\be \label{Tdual}
R \, \rightarrow \, 1/R
\ee
in the radius of the torus (in units of the string length $l_s$).
Under such a change, the energy spectrum of string states is
not modified if winding and momentum quantum numbers are interchanged
\be
(n, m) \, \rightarrow \, (m, n) \, .
\ee
The string vertex operators are consistent with this symmetry, and
thus T-duality is a symmetry of perturbative string theory. Postulating
that T-duality extends to non-perturbative string theory leads
\cite{Pol} to the need of adding D-branes to the list of fundamental
objects in string theory. With this addition, T-duality is expected
to be a symmetry of non-perturbative string theory.
Specifically, T-duality will take a spectrum of stable Type IIA branes
and map it into a corresponding spectrum of stable Type IIB branes
with identical masses \cite{Boehm}.

Since the number of string oscillatory modes increases exponentially
as the string mode energy increases, there is a maximal temperature
of a gas of strings in thermal equilibrium, the {\it Hagedorn
temperature} $T_H$ \cite{Hagedorn}. If we imagine taking a box of strings
and compressing it, the temperature will never exceed $T_H$. In fact,
as the radius $R$ decreases below the string radius, the temperature
will start to decrease, obeying the duality relation \cite{BV}
\be
T(R) \, = \, T(1/R) \, .
\ee
This argument shows that string theory has the potential of
taming singularities in physical observables. 
Similarly, the length $L$ measured by a physical observer will
be consistent with the symmetry (\ref{Tdual}), hence realizing
the idea of a minimal physical length \cite{BV}. Figure 2 provides
a sketch of how the temperature $T$ changes as a function of $R$.
\begin{figure}
\centering
\includegraphics[height=6cm]{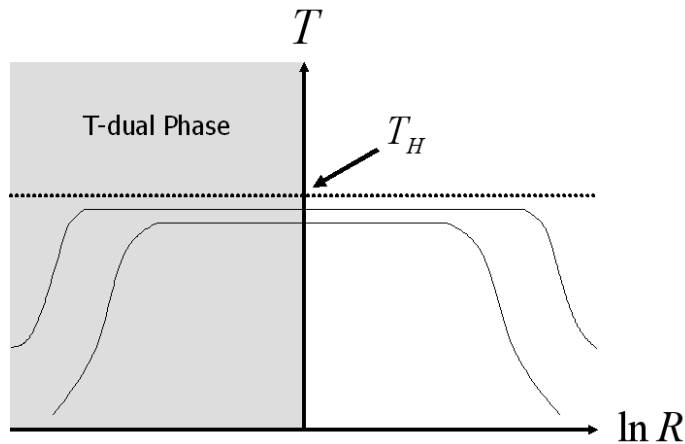}
\caption{Sketch (based on the analysis of \cite{BV}
of the evolution of temperature $T$ as a function
of the radius $R$ of space of a gas of strings in thermal
equilibrium. The top curve is characterized by an entropy
higher than the bottom curve, and leads to a longer
region of Hagedorn behaviour.}
\label{fig:2}       
\end{figure}

If we imagine that there is a dynamical principle that tells us how
$R$ evolves in time, then Figure 2 can be interpreted as depicting
how the temperature changes as a function of time. If $R$ is a monotonic
function of time, then two interesting possibilities for cosmology
emerge. If ${\rm ln}R$ decreases to zero at some fixed time (which without
loss of generality we can call $t = 0$),
and continues to decrease, we obtain a temperature profile which is
symmetric with respect to $t = 0$ and which (since small $R$ is physically
equivalent to large $R$) represents a bouncing cosmology 
(see \cite{Biswas2}
for a concrete recent realization of this scenario). If, on the other hand,
it takes an inifinite amount of time to reach $R = 0$, an {\it emergent
universe} scenario \cite{gellis} is realized.

It is important to realize that in both of the cosmological
scnearios which, as argued above, seem to follow from string theory symmetry
considerations alone, a large energy density does {\it not} lead
to rapid expansion in the Hagedorn phase, in spite of the fact that
the matter sources we are considering (namely a gas of strings) obeys all
of the usual energy conditions discussed e.g. in \cite{HE}). These
considerations are telling us that intuition drawn from Einstein gravity
may give us a completely incorrect picture of the early universe. This
provides a response to the main criticism raised in \cite{KKLM}.
 
Any physical theory requires both a specification of the equations of
motion and of the initial conditions.  We assume that
the universe starts out small and hot. For simplicity, we take
space to be toroidal, with radii in all spatial directions given by
the string scale. We assume that the initial energy density 
is very high, with an effective temperature which is close
to the Hagedorn temperature, the maximal temperature
of perturbative string theory. 

In this context, it was argued \cite{BV} that in order for spatial 
sections to become
large, the winding modes need to decay. This decay, at least
on a background with stable one cycles such as a torus, is only
possible if two winding modes meet and annihilate. Since string
world sheets have measure zero probability for intersecting in more
than four space-time dimensions, winding modes can annihilate only
in three spatial dimensions (see, however, the recent
caveats to this conclusion based on the work of \cite{Kabat3,Danos}). 
Thus, only three spatial dimensions
can become large, hence explaining the observed dimensionality of
space-time. As was shown later \cite{ABE}, adding branes to
the system does not change these conclusions since at later
times the strings dominate the cosmological dynamics.
Note that in the three dimensions which are becoming large there
is a natural mechanism of isotropization as long as some winding
modes persist \cite{Watson1}.

\subsection{Equations for String Gas Cosmology}

The above arguments were all heuristic. Some of them can be
put on a more firm mathematical basis, albeit in the context
of a toy model. The toy model consists of a classical
background coupled to a gas of strings. From the point of
view of rigorous string theory, this separation between
classical background and stringy matter is not satisfactory.
However, in the absence of a non-perturbative formulation of
string theory, at the present time we are forced to make this
separation. Note that this separation between classical
background geometry and string matter is common to all current
approaches to string cosmology.

As our background we choose dilaton gravity. It is crucial to include the
dilaton in the Lagrangian, firstly since
the dilaton arises in string perturbation theory at the same level
as the graviton (when calculating to leading order in the
string coupling and in $\alpha'$), 
and secondly because it is only the action of
dilaton gravity (rather than the action of Einstein gravity)
which is consistent with the T-duality symmetry. 
We will see, however, that the background dynamics inevitably drives
the system into a parameter region where the dilaton is
strongly coupled and hence beyond the region of validity
of the approximations made.
 
For the moment, however, we consider the dilaton gravity action 
coupled to a matter action $S_m$ 
\be
S \, = \, {1 \over {2 \kappa^2}} \int d^{10}x \sqrt{-{\hat g}} e^{-2 \phi}
\bigl[{\hat R} + 4 \partial^{\mu} \phi \partial_{\mu} \phi \bigr] + S_m \, ,
\ee
where ${\hat g}$ is the determinant of the metric, 
${\hat R}$ is the Ricci scalar,
$\phi$ is the dilaton, and
$\kappa$ is the reduced gravitational constant in ten dimensions.
The metric appearing in the above action is the metric in the
string frame. 

In the case of a homogeneous
and isotropic background given by
\be
ds^2 \, = \, dt^2 - a(t)^2 d{\bf x}^2 \, ,
\ee
the three resulting equations (the
generalization of the two Friedmann equations plus the equation
for the dilaton) in the string frame are
\cite{TV} (see also \cite{Ven})
\bea
-d {\dot \lambda}^2 + {\dot \varphi}^2 \, &=& \, e^{\varphi} E 
\label{E1} \\
{\ddot \lambda} - {\dot \varphi} {\dot \lambda} \, &=& \,
{1 \over 2} e^{\varphi} P \label{E2} \\
{\ddot \varphi} - d {\dot \lambda}^2 \, &=& \, {1 \over 2} e^{\varphi} E \, ,
\label{E3}
\eea
where $E$ and $P$ denote the total energy and pressure, respectively,
$d$ is the number of spatial dimensions, and we have introduced the
logarithm of the scale factor 
\be
\lambda(t) \, = \, {\rm log} (a(t))
\ee
and the rescaled dilaton
\be
\varphi \, = \, 2 \phi - d \lambda \, .
\ee

From the second of these equations it follows immediately
that a gas of strings containing both stable winding and
momentum modes will lead to the stabilization of the
radius of the torus: windings prevent expansion, momenta
prevent the contraction. The right hand side of the equation
can be interpreted as resulting from a confining potential for
the scale factor. 

Note that this behavior is a consequence
of having used dilaton gravity rather than Einstein gravity
as the background. The dilaton is evolving at the time when
the radius of the torus is at the minimum of its potential.
In fact, for the branch of solutions we are considering,
the dilaton is increasing as we go into the past. At
some point, therefore, it becomes greater than zero. At
this point, we enter the region of strong coupling. As
already discussed in \cite{Riotto}, a different dynamical
framework is required to analyze this phase. In
particular, the fundamental
strings are no longer the lightest degrees of freedom. We
will call this phase the ``strongly coupled Hagedorn phase"
for which we lack an analytical description. Since the
energy density in this phase is of the string scale, the
background equations should also be very different from
the dilaton gravity equations used above. In the following
section, we will make the assumption that the dilaton is
in fact frozen in the strongly coupled Hagedorn phase.
This could be a consequence of S-duality (see e.g. \cite{Kaya06}
for recent studies).

\subsection{String Gas Cosmology and Moduli Stabilization}

One of the outstanding issues when dealing with theories
with extra dimensions is the question of how the size and
shape moduli of the extra-dimensional spaces are stabilized.
String gas cosmology provides a simple and string-specific
mechanism to stabilize most of these moduli (see \cite{RHBrev6}
for a recent review). The outstanding issue is how to
stabilize the dilaton.

It is easiest to first understand radion stabilization in the
string frame \cite{Watson2}. The mechanism can be immediately
seen from the basic equations (\ref{E1} - \ref{E3}) of dilaton 
gravity. From (\ref{E2}) it follows that if the string gas
contains both winding and momentum modes, then there is a
preferred value for the scale factor. Winding modes prevent the
radion from increasing, momentum modes prevent it from decreasing.
If the number of winding and momentum modes is the same, then
the fixed point of the dynamics corresponds to string-scale
radion. As can be seen from (\ref{E3}), the dilaton
is evolving when the radion is at the fixed point.

In order to make contact with late-time cosmology, we need to
assume that the dilaton is fixed by some non-perturbative
mechanism. The issue of radion stabilization is then no longer
this simple. A string matter state which minimizes the
effective potential in the Einstein frame is only consistent
with stabilization if the energy density vanishes at the
mininum - otherwise, the state will in fact lead to
inflationary expansion. It thus turns out that \cite{Subodh1,Subodh2} (see
also \cite{Watson2,BattWat2}) states which are massless and
have an enhanced symmetry at a particular radius play a crucial role. 
If such states exist, then
the radion can be fixed at this particular radius (in our case
it will be the self-dual radius). Such states
exist in the heterotic string theory, but not in Type II string
theories where they are projected out by the GSO projection.

To understand stabilization in the Einstein frame,
let us consider the equations of motion which arise from coupling
the Einstein action (as opposed to the dilaton gravity action)
to a string gas. In the anisotropic setting when the metric
is taken to be
\be
ds^2 \, = \, dt^2 - a(t)^2 d{\bf x}^2 - 
\sum_{\alpha = 1}^6 b_{\alpha}(t)^2 dy_{\alpha}^2 \, ,
\ee
where the $y_{\alpha}$ are the internal coordinates, the
equations of motion for $b_{\alpha}$ becomes
\be \label{extra}
{\ddot b_{\alpha}} + 
\bigl( 3 H + \sum_{\beta = 1, \beta \neq \alpha}^6 {{{\dot b_{\beta}}} \over {b_{\beta}}} \bigr) {\dot b_{\alpha}} \, = \, 
\sum_{n, m} 8 \pi G {{\mu_{m,n}} \over {\sqrt{g} \epsilon_{m,n}}}{\cal S} \,
\ee
where $\mu_{m,n}$ is the number density of $(m,n)$ strings, 
$\epsilon_{m,n}$ is the energy of an individual $(m,n)$ string, 
and $g$ is the determinant of the metric. The source term 
${\cal S}$ depends on the quantum numbers of the
string gas, and the sum runs over all momentum numbers and winding
number vectors $m$ and $n$, respectively (note that $n$ and $m$ are
six-vectors, one component for each internal dimension). If the number
of right-moving oscillator modes is given by $N$, then the source term
for fixed $m$ and $n$ is
\be \label{source}
{\cal S} \, = \, \sum_{\alpha} \bigl( {{m_{\alpha}} \over {b_{\alpha}}} \bigr)^2
- \sum_{\alpha} n_{\alpha}^2 b_{\alpha}^2 
+ {2 \over {D - 1}} \bigl[ (n,n) + (n, m) + 2(N - 1) \bigr] \, .
\ee
To obtain this equation, we have made use of the mass spectrum of
string states and of the level matching conditions. In the case of
the bosonic superstring, the mass spectrum for fixed $m, n, N$ and 
${\tilde N}$,
where ${\tilde N}$ is the number of left-moving oscillator states,
on a six-dimensional torus whose radii are given by $b_{\alpha}$ is
\be
m^2 \, = \, \bigl({{m_{\alpha}} \over {b_{\alpha}}} \bigr)^2
- \sum_{\alpha} n_{\alpha}^2 b_{\alpha}^2 + 2 (N + {\tilde N} - 2) \, ,
\ee
and the level matching condition reads
\be
{\tilde N} \, = \, (n,m) + N \, ,
\ee
where $(n,m)$ indicates the scalar product of $n$ and $m$ in the 
trivial internal metric.

There are modes which are massless at the self-dual radius 
$b_{\alpha} = 1$.
One such mode is the graviton with $n = m = 0$ and $N = 1$. The modes of
interest to us are modes which contain winding and momentum, namely 
\begin{itemize}
\item{} $N = 1$, $(m,m) = 1$, $(m, n) = -1$ and $(n,n) = 1$;
\item{} $N = 0$, $(m,m) = 1$, $(m, n) = 1$ and $(n,n) =  1$;
\item{} $N = 0$  $(m,m) = 2$, $(m, n) = 0$ and $(n,n) =  2$.
\end{itemize}
Note that some of these modes survive in the Heterotic string theory, but
they do not survive the GSO \cite{Pol} truncation in Type II string
theories.

In string theories which admit massless states (i.e. states
which are massless at the self-dual radius), these states
will dominate the initial partition function. The background
dynamics will then also be dominated by these states. To understand
the effect of these strings, consider the equation of motion (\ref{extra})
with the source term (\ref{source}). The first two terms in the
source term correspond to an effective potential with a stable
minimum at the self-dual radius. However, if the third term in the
source does not vanish at the self-dual radius, it will lead to
a positive potential which causes the radion to increase. Thus,
a condition for the stabilization of $b_{\alpha}$ at the self-dual
radius is that the third term in (\ref{source}) vanish at the
self-dual radius. This is the case if and only if the string state
is a massless mode.

The massless modes have other nice features which are explored in
detail in \cite{Subodh2}. They act as radiation from the
point of view of our three large dimensions and hence do not
lead to an over-abundance problem. As our three spatial dimensions
grow, the potential which confines the radion becomes shallower.
However, rather surprisingly, it turns out the the potential
remains steep enough to avoid fifth force constraints. 

Key to the success in simultaneously avoiding the moduli overclosure
problem and evading fifth force constraints is the fact that
the stabilization mechanism is an intrinsically stringy one,
as opposed to an effective field theory mechanism. In the case
of effective field theory, both the confining force and the overdensity
in the moduli field scale as $V(\varphi)$, where $V(\varphi)$ is the
potential energy density of the field $\varphi$. In contrast, in
the case of stabilization by means of massless string modes, the
energy density in the string modes (from the point of view of
our three large dimensions) scales as $p_3$, whereas the confining
force scales as $p_3^{-1}$, where $p_3$ is the momentum in the three
large dimensions. Thus, for small values of $p_3$, one simultaneously
gets large confining force (thus satisfying the fifth force constraints) 
and small energy density \cite{Subodh2}.

In the presence of massless string states, the shape moduli also
can be stabilized, at least in the simple toroidal backgrounds
considered so far \cite{Edna} (see also \cite{Sugumi}). The
stabilization mechanism is once again dynamical, and makes use
of the massless states with enhanced symmetry (see also
\cite{Watson3,Eva} for more on the use of massless states with
enhanced symmetries in cosmology).

There has been other recent work on string gas cosmology. The
interested reader is referred to \cite{other}. In particular,
the difficulties of dilaton stabilization have been addressed
in \cite{dilaton}. For recent attempts to stabilize the dilaton
see \cite{Subodh3} and \cite{Kaya06}.

\section{String Gas Cosmology and Structure Formation}

\subsection{Overview}

Let us recall the key aspects of the dynamics of string gas
cosmology. First of all, note that in thermal equilibrium at the 
string scale ($R \simeq l_s$), the
self-dual radius, the number of winding and momentum modes will be 
equal. Since winding and momentum modes give an opposite 
contribution to the pressure, the pressure of the string gas in 
thermal equilibrium at the self-dual radius will vanish. From  
the dilaton gravity equations of motion (\ref{E1} - \ref{E3})
it then follows that a static phase 
$\lambda = 0$ will be a fixed point of the dynamical system. This 
phase is the Hagedorn phase. 
 
On the other hand, for large values of $R$ in thermal equilibrium 
the energy will be exclusively in momentum modes. These act as usual 
radiation. Inserting the radiative equation of state into the above 
equations (\ref{E1} - \ref{E3}) it follows that the source in 
the dilaton equation of motion vanishes and the dilaton approaches 
a constant as a consequence of the Hubble damping term in its
equation of motion. Consequently, the 
scale factor expands as in the usual radiation-dominated universe.

The transition between the Hagedorn phase and the radiation-dominated 
phase with fixed dilaton is achieved via the annihilation of winding 
modes, as studied in detail in \cite{BEK}. 
The main point is that, 
starting in a Hagedorn phase, there will be a smooth transition to 
the radiation-dominated phase of standard cosmology with fixed dilaton. 

Our new cosmological background is obtained by following our
currently observed universe into the past according to the
string gas cosmology equations. The radiation phase of standard
cosmology is unchanged. In particular, the dilaton is fixed
in this phase \footnote{The dilaton comes to rest, but it is
not pinned to a particular value by a potential. Thus, in order
to obtain consistency with late time cosmology, an additional
mechanism operative at late times which fixes the dilaton is
required.}. However, as the temperature of the radiation
bath approaches the Hagedorn temperature, the equation of state
of string gas matter changes. The equation of state parameter
$w = P / E$ decreases towards a pressureless state and the
string frame metric becomes static. Note that, in order
for the present size of the universe to be larger than our current
Hubble radius, the size of the spatial sections in the
Hagedorn phase must be at least $1 {\rm mm}$
\footnote{How to obtain this initial size starting from string-scale
initial conditions constitutes the {\it entropy problem} of
our scenario. A possible solution making use of an initial phase of
bulk dynamics is given in \cite{Natalia}.}. We will denote
the time when the transition from the Hagedorn phase to the radiation
phase of standard cosmology occurs by $t_R$, to evoque the analogy
with the time of reheating in inflationary cosmology. As we go back
in time in the Hagedorn phase, the dilaton increases. At the time
$t_c$ when the dilaton equals zero, a second transition occurs, the
transition to a ``strongly coupled Hagedorn phase'' (using the
terminology introduced in \cite{Betal}). We take the dilaton to be
fixed in this phase. In this case, the strongly coupled Hagedorn
phase may have a duration which is very long compared to the Hubble
time immediately following $t_R$. 
It is in this cosmological background that we will study the 
generation of fluctuations.  

It is instructive to compare the background evolution of string gas 
cosmology with the background of inflationary cosmology. 
Figure 3 is a sketch of 
the space-time evolution in string gas cosmology. For times $t < t_R$, 
we are in the quasi-static Hagedorn phase, for $t > t_R$ we have 
the radiation-dominated period of standard cosmology. To understand 
why string gas cosmology can lead to a causal mechanism of structure 
formation, we must compare the evolution of the physical 
wavelength corresponding to a fixed comoving scale (fluctuations in 
early universe cosmology correspond to waves with a fixed wavelength 
in comoving coordinates) with that of the Hubble radius $H^{-1}(t)$, 
where $H(t)$ is the expansion rate. The Hubble radius separates 
scales on which fluctuations oscillate (wavelengths smaller than 
the Hubble radius) from wavelengths where the fluctuations are frozen 
in and cannot be effected by microphysics. Causal microphysical 
processes can generate fluctuations only on sub-Hubble scales 
(see e.g. \cite{RHBrev2} for 
a concise overview of the theory of cosmological perturbations and 
\cite{MFB} for a comprehensive review). 
 
\begin{figure} 
\includegraphics[height=10cm]{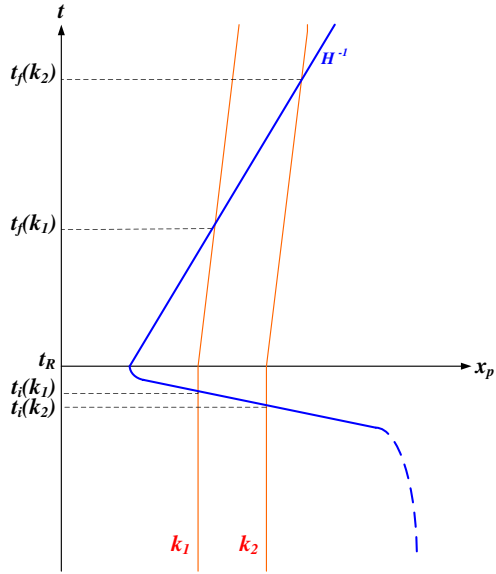} 
\caption{Space-time diagram (sketch) showing the evolution of fixed 
comoving scales in string gas cosmology. The vertical axis is time, 
the horizontal axis is physical distance. The Hagedorn phase ends at 
the time $t_R$ and is followed by the radiation-dominated phase of 
standard cosmology. The solid curve represents the Hubble radius 
$H^{-1}$ which is cosmological during the quasi-static Hagedorn 
phase, shrinks abruptly to a microphysical scale at $t_R$ and then 
increases linearly in time for $t > t_R$. Fixed comoving scales (the 
dotted lines labeled by $k_1$ and $k_2$) which are currently probed 
in cosmological observations have wavelengths which are smaller than 
the Hubble radius during the Hagedorn phase. They exit the Hubble 
radius at times $t_i(k)$ just prior to $t_R$, and propagate with a 
wavelength larger than the Hubble radius until they reenter the 
Hubble radius at times $t_f(k)$.} \label{fig:3} 
\end{figure}

In string gas cosmology, the Hubble radius is infinitely large in the 
Hagedorn phase. As the universe starts expanding near $t_R$, the 
Hubble radius rapidly decreases to a microscopic value (set by the 
temperature at $t = t_R$ which will be close to the Hagedorn 
temperature), before turning around and increasing linearly in the 
post-Hagedorn phase. The physical wavelength corresponding to a 
fixed comoving scale, on the other hand, is constant during the 
Hagedorn era. Thus, all scales on which current experiments measure 
fluctuations are sub-Hubble deep in the Hagedorn phase. In the 
radiation period, the physical wavelength of a perturbation mode 
grows only as $\sqrt{t}$. Thus, at a late time $t_{f}(k)$ the 
fluctuation mode will re-enter the Hubble radius, leading to the 
perturbations which are observed today. 
 
In contrast, in inflationary cosmology (Figure 1) the Hubble radius 
is constant during inflation ($t < t_R$, where here $t_R$ is the 
time of inflationary reheating), whereas the physical wavelength 
corresponding to a fixed comoving scale expands exponentially. Thus, 
as long as the period of inflation is sufficiently long, all scales 
of interest for current cosmological observations are sub-Hubble at 
the beginning of inflation. 
 
In spite of the fact that both in inflationary cosmology and in string 
gas cosmology, scales are sub-Hubble during the early stages, the 
actual generation mechanism for fluctuations is completely different. 
In inflationary cosmology, any thermal fluctuations present before 
the onset of inflation are red-shifted away, leaving us with a quantum 
vacuum state, whereas in the quasi-static Hagedorn phase of string gas 
cosmology matter is in a thermal state. Hence, whereas in inflationary 
cosmology the fluctuations originate as quantum vacuum perturbations, 
in string gas cosmology the inhomogeneities are created by the thermal 
fluctuations of the string gas. 
 
As we have shown in \cite{NBV,Ali,BNPV2}, string thermodynamical
fluctuations in 
the Hagedorn phase of string gas cosmology yield an almost 
scale-invariant spectrum of both scalar and tensor modes. This
result stems from the holographic scaling of the specific heat
$C_V(R)$ (evaluated for fixed volume) as a function of the radius
$R$ of the box
\be \label{specheat}
C_V(R) \, \sim \, R^2 \, .
\ee
As derived in \cite{Deo}, this result holds true for a gas of closed
strings in a space-time in which the three large spatial dimensions
are compact.
The scaling (\ref{specheat}) is an 
intrinsically stringy result: thermal fluctuations of a gas of 
particles would lead to a very different  
scaling. 

Since the primordial perturbations in our scenario are of 
thermal origin (and there are no non-vanishing chemical potentials), 
they will be adiabatic. The spectrum of scalar metric fluctuations
has a slight red tilt. As a distinctive feature \cite{BNPV1}, our
scenario predicts a slight blue tilt for the spectrum of gravitational
waves. The red tilt for the scalar modes is due to the fact that the
temperature when short wavelength modes exit the Hubble radius is
slightly lower than the temperature when longer wavelength modes exit.
The gravitational wave amplitude, in contrast, is determined by
the pressure. Since the pressure is closer to zero the deeper in the
Hagedorn phase we are, a slight blue tilt for the tensor fluctuations
results.

These results are explained in more detail in the following
subsections. 
   
\subsection{Extracting Metric Fluctuations from Matter Perturbations}

In this subsection, we show how the scalar and tensor metric fluctuations 
can be extracted from knowledge of the energy-momentum tensor of 
the string gas. 
 
Working in conformal time $\eta$ (defined via 
$dt = a(t)d\eta$), the metric of a homogeneous and isotropic background 
space-time perturbed by linear scalar metric fluctuations and 
gravitational waves can be written in the form 
\be \label{pertmetric}
d s^2 = a^2(\eta) \left\{(1 + 2 \Phi)d\eta^2 - [(1 - 
2 \Phi)\delta_{ij} + h_{ij}]d x^i d x^j\right\} \,. 
\ee 
Here, $\Phi$ (which is a function of space and time) describes 
the scalar metric fluctuations. The tensor $h_{ij}$ is 
transverse and traceless and contains the two polarization states of 
the gravitational waves. In the above, we have fixed the coordinate 
freedom by working in the so-called ``longitudinal" gauge in which 
the scalar metric fluctuation is diagonal. We have also assumed
that there is no anisotropic stress. Note that to linear order 
in the amplitude of the fluctuations, scalar and tensor modes 
decouple, and the tensor modes are gauge-invariant.

Our approximation scheme for computing the cosmological 
perturbations and gravitational wave spectra from string gas 
cosmology is as follows (the analysis is similar to how 
the calculations were performed in \cite{BST,BK} in the 
case of inflationary cosmology). For a fixed comoving scale $k$ we follow 
the matter fluctuations until the time $t_i(k)$ 
shortly before the end of the Hagedorn phase when the scale exits 
the Hubble radius \footnote{Recall that on sub-Hubble scales, the
dynamics of matter is the dominant factor in the evolution of the
system, whereas on super-Hubble scales, matter fluctuations freeze
out and gravity dominates. Thus, it is precisely at the
time of Hubble radius crossing that we must extract the
metric fluctuations from the matter perturbations. Since the
concept of an energy density fluctuation is gauge-dependent on
super-Hubble scales, one cannot extrapolate the matter spectra
to larger scales as was suggested in Section 3 of \cite{KKLM}.} 
At the time of Hubble radius crossing, we use the Einstein constraint
equations (discussed below) to compute the values of 
$\Phi(k)$ and $h(k)$ ($h$ is 
the amplitude of the gravitational wave tensor), and then we
propagate the metric fluctuations according to the standard
gravitational perturbation equations until scales re-enter the
Hubble radius at late times. Note that since the dilaton is
fixed in the radiation phase, we are justified in using the
perturbed Einstein equations after the time $t_R$.

Since the dilaton comes to rest at the end of the Hagedorn phase,
we will use the Einstein equations to relate the matter fluctuations
to the metric perturbations at the time that the scales $k$ exit the
Hubble radius at times $t_i(k)$. Since $t_i(k) < t_R$, there are
potentially important terms coming from the dilaton velocity
which we are neglecting \cite{Betal,KKLM}. We will return to this issue later.
 
Inserting the metric (\ref{pertmetric}) into the Einstein equations, 
subtracting the background terms and truncating the perturbative expansion 
at linear order leads to the following system of equations 
\begin{eqnarray} \label{perteom1} 
- 3 {\cal H} \left( {\cal H} \Phi + \Phi^{'} \right) + \nabla^2 \Phi 
\, 
&=& \, 4 \pi G a^2 \delta T^0{}_0 \nonumber \\ 
\left( {\cal H} \Phi + \Phi^{'} \right)_{, i} \, 
&=& 4 \pi G a^2 \delta T^0{}_i  \nonumber \\ 
\left[ \left( 2 {\cal H}^{'} + {\cal H}^2 \right) \Phi + 3 {\cal H} 
\Phi^{'} 
+ \Phi^{''} \right] 
&=& - 4 \pi G a^2 \delta T^i{}_i \, , \nonumber \\ 
-{1 \over 2} \left[ {{\cal H}^{\prime}} + {1 \over 
2}{\cal H}^2 \right] h_{ij} 
+ {1 \over 4} {\cal H} h_{ij}^{\prime} && \nonumber \\ 
+ \left[{{\partial^2} \over {\partial \eta^2}} - \nabla^2\right] 
h_{ij} \, 
&=& - 4 \pi G a^2 \delta T^i{}_j \,,\nonumber \\ 
&& \mbox{for $i \neq j$}\,, 
\end{eqnarray} 
where ${\cal H} = a^{\prime} / a$, a prime 
denotes the derivative with respect to conformal time $\eta$, 
and $G$ is Newton's gravitational constant. 
 
In the Hagedorn phase, these equations simplify substantially and 
allow us to extract the scalar and tensor metric fluctuations 
individually. Replacing comoving by 
physical coordinates, we obtain from the $00$ equation 
\be 
\label{scalar} \nabla^2 \Phi \, = \, 4 \pi G \delta T^0{}_0 
\, 
\ee 
and from the $i \neq j$ equation 
\be 
\label{tensor} \nabla^2 h_{ij} \, = \, - 4 \pi G \delta T^i{}_j 
\, . 
\ee 

The above equations (\ref{scalar}) and (\ref{tensor}) allow 
us to compute the power spectra of scalar and tensor metric 
fluctuations in terms of correlation functions of the string
energy-momentum tensor. Since the metric perturbations are 
small in amplitude we can consistently work in Fourier space. 
Specifically, 
\be \label{scalarexp} 
\langle|\Phi(k)|^2\rangle \, = \, 16 \pi^2 G^2 
k^{-4} \langle\delta T^0{}_0(k) \delta T^0{}_0(k)\rangle \, , 
\ee 
where the pointed brackets indicate expectation values, and 
\be 
\label{tensorexp} \langle|h(k)|^2\rangle \, = \, 16 \pi^2 G^2 
k^{-4} \langle\delta T^i{}_j(k) \delta T^i{}_j(k)\rangle \,, 
\ee 
where on the right hand side of (\ref{tensorexp}) we mean the 
average over the correlation functions with $i \neq j$. 

\subsection{String Thermodynamics Fluctuations}

Since the Hagedorn phase is quasi-static and dominated by a
gas of strings, fluctuations in our scenario are the thermal
fluctuations of a gas of strings. We will consider a gas of
closed strings in a compact space, i.e. our three-dimensional
space is considered to be large but compact. Specifically, it is
important to have winding modes in the spectrum of string states.

The correlation functions of the energy-momentum tensor can
be obtained from the partition function $Z$, which 
determines the free energy $F$ via
\be \label{free} 
F \, = \, \frac{-1}{\beta} ln Z \, , 
\ee 
where $\beta$ is the inverse temperature.

The expectation value of the energy-momentum tensor $T_{\mu}^{\nu}$
is then given in terms of the free energy by
\be \label{texp2} 
\langle T_\nu^\mu \rangle \, = \, \frac{-2 g_{\nu \lambda}}{\sqrt{-g}} 
\frac{\delta F}{\delta g_{\lambda \mu}} \, .   
\ee 
Taking one additional variational derivative of (\ref{texp2}) we obtain 
the following expression for the fluctuations of $T_{\mu}^{\nu}$
(see \cite{Ali,BNPV2} for more details):
\be 
\label{fluct} 
\langle T^\mu_\nu T^\kappa_\beta \rangle -  
\langle T^\mu_\nu \rangle \langle T^\kappa_\beta \rangle 
\,\, = \frac{1}{\beta}\frac{2 g_{\beta \tau}}{\sqrt{-g}} 
\frac{\delta }{\delta g_{\tau \kappa}}\Bigl( 
\frac{-2 g_{\nu \lambda}}{\sqrt{-g}}\frac{\delta F }{\delta g_{\lambda \mu}}  
\Bigr) \, . 
\ee
In particular, the energy density fluctuation is
\be \label{cor1}
\langle \delta\rho^2 \rangle \, = \,  
- \frac{1}{R^{6}} \frac{\partial}{\partial \beta} 
\left(F + \beta \frac{\partial F}{\partial \beta}\right) \, = \, 
\frac{T^2}{R^6} C_V \, , 
\ee 
where $C_V$ is the specific heat. The off-diagonal pressure
fluctuations, in turn, are given by
\bea \label{cor2} 
\langle \delta {T^i{}_j}^2 
\rangle \, &=& \, \langle {T^i{}_j}^2 \rangle - \langle T^i{}_j \rangle^2 \\ 
&=& \, \frac{1}{\beta R^3}\frac{\partial}{\partial \ln{R}}\left(- 
\frac{1}{R^3} \frac{\partial F}{\partial \ln{R}}\right)  = 
\frac{1}{\beta R^2}\frac{\partial p}{\partial R} \, , \nonumber
\eea 
where the string pressure is given by 
\be 
p \,  \equiv   \, - V^{-1}(\partial F/\partial \ln{R}) \, = \, T 
(\partial S/\partial V)_E  \, . \label{stringypressure} 
\ee 

So far, the analysis has been general thermodynamics. Let us
now specialize to the thermodynamics of strings.
In \cite{Deo}, the thermodynamical properties of a gas of closed
strings in a toroidal space of radius $R$ were computed. 
To compute the fluctuations in a region 
of radius $R$ which forms part of our three-dimensional compact
space, we will apply the results of \cite{Deo} for a box of
strings in a volume $V = R^3$. 

The starting point of the computation is the formula for 
the density of states $\Omega(E,R)$ which determines the entropy $S$ via
\be 
S(E , R) \, = \, \ln{\Omega(E ,R)} \, .
\ee 
The entropy, in turn, determines the free energy $F$ from which
the correlation functions can be derived. In the Hagedorn phase,
the density of states has the following form
\be \label{entropy}
\Omega(E , R) \, \simeq \, \beta_H e^{\beta_H E + n_H V}[1 + 
\delta \Omega_{(1)}(E , R)] \label{density_states}\,, 
\ee 
where $\beta_H = T_H^{-1}$, $n_H$ is a (constant) number density of order 
$\ell_s^{-3}$ ($\ell_s$ being the string length), $\rho_H$ is the 
`Hagedorn Energy density' of the order $\ell_s^{-4}$, and
\be 
\label{deltaomega} 
\ell_s^3\delta \Omega_{(1)} \, \simeq \, - \frac{R^2}{T_H}\left(1 - 
\frac{T}{T_H}\right) \, . 
\ee 
In addition, we find 
\be \label{meanen} 
\langle E \rangle \, \simeq \,  \ell_s^{-3} R^2 
\ln{\left[\frac{\ell_s^3 T}{R^2 (1- T/T_H)}\right]}\,. 
\ee 
Note that to ensure that $|\delta \Omega_{(1)}| \ll 1$ and $\langle E 
\rangle  \gg \rho_H R^3$, one should demand  
\be \label{cond} 
(1 - T/T_H) R^2 \ell_s^{-2} \ll 1 \, . 
\ee 
 
The results (\ref{entropy}) and (\ref{deltaomega}) now allow us 
to compute the correlation functions (\ref{cor1}) and (\ref{cor2}). 
We first compute the energy correlation function (\ref{cor1}). 
Making use of (\ref{meanen}), it follows from 
(\ref{specheat}) that 
\be \label{specheat2} 
C_V  \, \approx \, 2 \frac{R^2/\ell^3}{T \left(1 - T/T_H\right)}\, . 
\ee 
The `holographic' scaling $C_V(R) \sim R^2$ is responsible for the
overall scale-invariance of the spectrum of cosmological perturbations. 
The factor $(1 - T/T_H)$ in the denominator is responsible 
for giving the spectrum a slight red tilt. It comes from the 
differentiation with respect to $T$. 

For the pressure. we obtain 
\be 
p(E, R) \approx n_H T_H - \frac{2}{3}\frac{(1 - T/T_H)}{\ell_s^3 
R}\ln{\left[\frac{\ell_s^3 T}{R^2 (1- T/T_H)}\right]} \,, 
\ee 
which immediately yields 
\be \label{tensorresult} 
\langle \delta {T^i{}_j}^2 \rangle \, 
\simeq \, \frac{T (1 - T/T_H)}{\ell_s^3 R^4} 
\ln^2{\left[\frac{R^2}{\ell_s^2}(1 - T/T_H)\right]}\, .  
\ee 
Note that since no temperature derivative is taken, the factor $(1 - T/T_H)$ 
remains in the numerator. This leads to the 
slight blue tilt of the spectrum of gravitational waves. As mentioned
earlier, the physical reason for this blue tilt is that larger wavelength
modes exit the Hubble radius deeper in the Hagedorn phase where the
pressure is smaller and thus the strength of the tensor modes is less.

\subsection{Power Spectra}

The power spectrum of scalar metric fluctuations is given by
\bea \label{power2} 
P_{\Phi}(k) \, & \equiv & \, k^3 |\Phi(k)|^2 \\
&=& \, 16 \pi^2 G^2 k^{-1} <|\delta \rho(k)|^2> \, . \nonumber \\
&=& \, 16 \pi^2 G^2 k^2 <(\delta M)^2>_R \nonumber \\ 
               &=& \, 16 \pi^2 G^2 k^{-4} <(\delta \rho)^2>_R 
\, , \nonumber 
\eea 
where in the first step we have used (\ref{scalarexp}) to replace the 
expectation value of $|\Phi(k)|^2$ in terms of the correlation function 
of the energy density, and in the second step we have made the 
transition to position space (note that $k = R^{-1}$).

According to (\ref{cor1}), the density correlation function 
is given by the specific heat via $T^2 R^{-6} C_V$.
Inserting the expression from (\ref{specheat2}) for the specific 
heat of a string gas on a scale $R$ yields to the final result 
\be \label{power4} 
P_{\Phi}(k) \, = \, 16 \pi^2 G^2 {T \over {\ell_s^3}} {1 \over {1 - T/T_H}} 
\ee 
for the power spectrum of cosmological fluctuations. In the above
equation, the temperature $T$ is to be evaluated at the time $t_i(k)$
when the mode $k$ exits the Hubble radius. Since modes with larger
values of $k$ exit the Hubble radius slightly later when the
temperature is slightly lower, a small red tilt of the spectrum is
induced. The amplitude ${\cal A}_S$ of the power spectrum is 
given by
\be
{\cal A}_S \, \sim \, \bigl({{l_{pl}} \over {l_s}}\bigr)^4 
{1 \over {1 - T/T_H}} \, .
\ee
Taking the last factor to be of order unity, we find that a string
length three orders of magnitude larger than the Planck length,
a string length which was assumed in early studies of string theory,
gives the correct amplitude of the spectrum. Thus, it appears that
the string gas cosmology structure formation mechanism does not have
a serious amplitude problem.

Similarly, we can compute the power spectrum of the gravitational waves
and obtain 
\be \label{tpower3} 
P_h(k) \, \sim \, 16 \pi^2 G^2 {T \over {\ell_s^3}} (1 - 
T/T_H) \ln^2{\left[\frac{1}{\ell_s^2 k^2}(1 - T/T_H)^{-1}\right]}\, .  
\ee 
This shows that the spectrum of tensor modes is - to a first approximation, 
namely neglecting the logarithmic factor and neglecting the k-dependence 
of $T(t_i(k))$ - scale-invariant \footnote{We believe that the calculation
of Section 4 in \cite{KKLM} which yields a result with different slope and
much smaller amplitude is based on a temporal
Green function calculation which misses the initial condition
term which dominates the spectrum.}. The k-dependence of the temperature
at Hubble radius crossing induces a small blue tilt for the spectrum
of gravitational waves.

Comparing (\ref{power4}) and (\ref{tpower3}) we see that the tensor
to scalar ratio is suppressed by the factor $(1 - T/T_H)^2$.
Given a good understanding of the exit from the Hagedorn phase we
would be able to compute this ratio as well as the magnitude of the
spectral tilts for both scalar and tensor modes.

\subsection{The Strongly Coupled Hagedorn Phase} 
 
In the previous discussion we have assumed that during the Hagedorn
phase, the kinetic energy of the dilaton has negligible effects.
However, this is not the case \cite{Betal,KKLM} if the background is
described in terms of dilaton gravity. However, as stressed in Section 2.6,
it is unrealisitc to expect that pure dilaton gravity is a good
approximation to the dyanamics of the Hagedorn phase. Both string
loop and $\alpha^{\prime}$ corrections will be important. Another aspect
of this issue is that - according to the dilaton gravity equations -
the string coupling constant quickly becomes greater than unity as we
go back in time from $t_R$. At that point, the separation between
classical dilaton background and stringy matter becomes untenable.

Therefore, in order to put our scenario on a firm basis, we need a
consistent description of the Hagedorn phase. It is crucial that there
be a phase before or immediately leading up to $t_R$, the time when
the winding string modes decay to string loops, when the dilaton is
fixed. In this case, the calculations we have done above are well
justified.

Let us assume, for example, that the dilaton gets fixed once it reaches
its self-dual value, at a time which we denote by $t_c$. Freezing
the dilaton allows the Hagedorn phase to be of sufficiently long
duration for thermal string equilibrium to be established on scales
up to $1 {\rm mm}$. It will also put out structure formation
scenario on a more solid footing. In this case,
the space-time diagram is given by Fig. 4, where we are now plotting
scales consistently in the Einstein frame. As is apparent, the overall
structure of the diagram is the same as that of Fig. 3, except for
the fact that scales exit the Einstein frame Hubble radius at a time
$t_i(k)$ immediately before $t_c$ instead of immediately before $t_R$.
A valid concern is that we might not be allowed to neglect the higher
derivative terms in the gravitational action at such early times during
the Hagedorn phase.

\begin{figure}
\includegraphics[height=6cm]{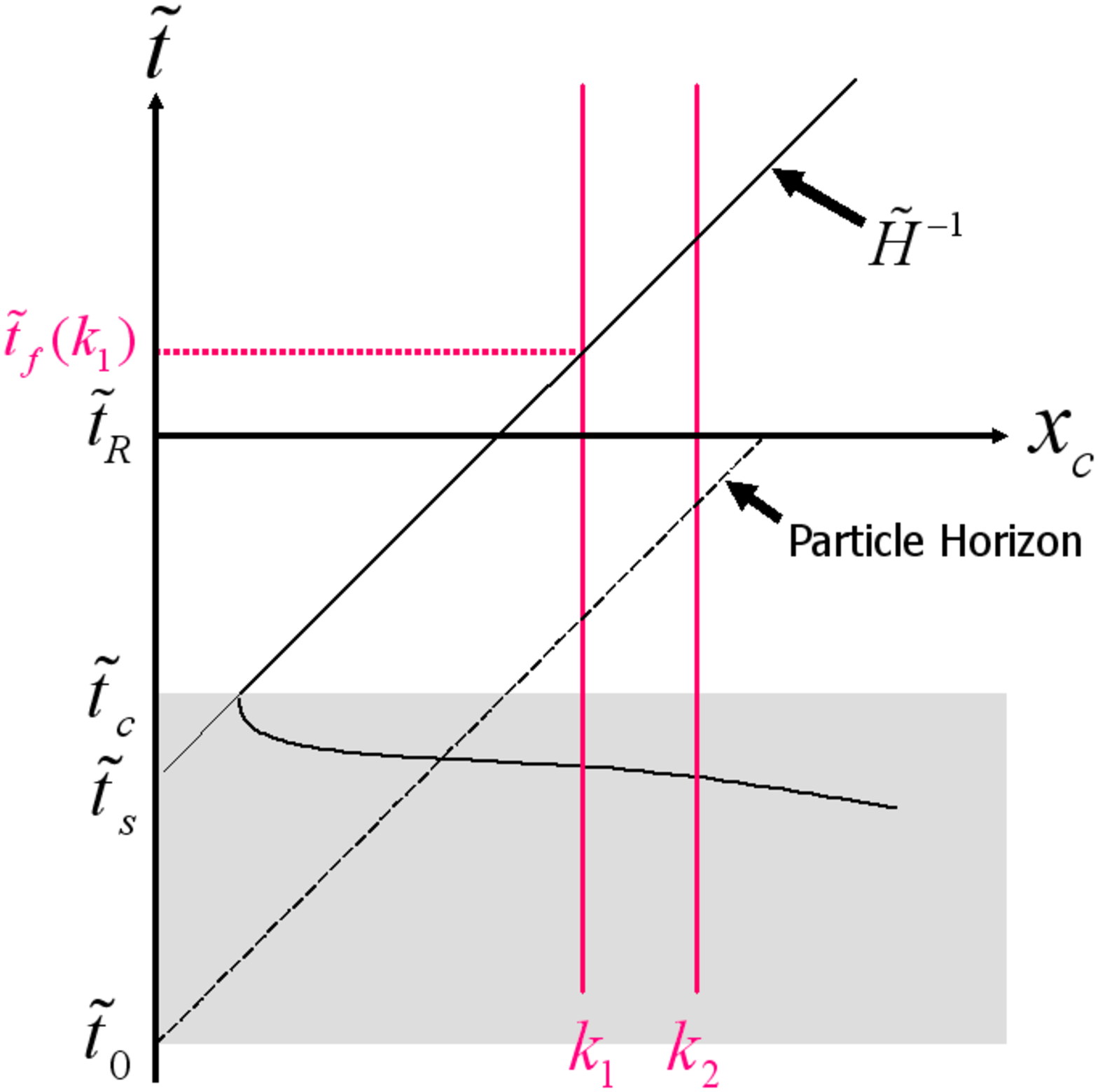}
\begin{caption}
{\small Space-time diagram (sketch) showing the evolution
of fixed comoving scales in string gas cosmology. 
The vertical axis is Einstein frame time, the horizontal
axis is comoving distance.  The solid curve represents the 
Einstein frame Hubble radius ${\tilde{H}}^{-1}$ which is linearly
increasing after ${\tilde{t}}_c$. Fixed comoving scales
(the dotted lines labeled by $k_1$ and $k_2$) which are currently probed in
cosmological observations have wavelengths which are larger than
the Einstein frame Hubble radius during the part of the Hagedorn phase
in which the dilaton is rolling. However, due to the presence of
the initial strong coupling Hagedorn phase, the horizon becomes much 
larger than the Hubble radius. The shaded region corresponds to the 
strong coupling Hagedorn phase.}
\end{caption}
\label{fig:4}
\end{figure}

A background in which our string gas structure formation scenario can
be implemented \cite{Biswas2} is the ghost-free and asymptotically
free higher derivative gravity model proposed in \cite{Biswas} given
by the gravitational action
\be
S \, = \, \int d^4x \sqrt{-g} F(R)
\ee
with
\be
F(R) \, = \, R + \sum_{n = 0}^{\infty} {{c_n} \over {M_s^{2n}}} R 
\bigl( {{\partial^2} \over {\partial^2 t}} - \nabla^2 \bigr)^n R
\, ,
\ee
where $M_s$ is the string mass scale (more generally, it is the scale
where non-perturbative effects start to dominate), and the $c_n$ are
coefficients of order unity.

As shown in \cite{Biswas} and \cite{Biswas2}, this action has bouncing
cosmological solutions. If the temperature during the bounce phase is
sufficiently high, then a gas of strings will be excited in this
phase. In the absence of initial cosmological perturbations in the
contracting phase, our string gas structure formation scenario is
realized. The string network will contain winding modes in the same
way that a string network formed during a cosmological phase transition
will contain infinite strings. The dilaton is fixed in this scenario,
thus putting the calculation of the cosmological perturbations on
a firm basis. There are no additional dynamical degrees of freedom
compared to those in Einstein gravity. The higher derivative
corrections to the equations of motion (in particular to the
Poisson equation) are suppressed by factors of $(k/M_s)^2$. Thus,
all of the conditions on a cosmological backgroound to successfully
realize the string gas cosmology structure formation scenario are
realized.

\section{Discussion}

In spite of the great phenomenological successes at solving key
problems of Standard Big Bang cosmology such as the horizon, entropy
and flatness problems, and at providing a simple and predictive
structure formation scenario, current realizations of inflation
are beset by important conceptual problems. Most importantly,
the applicability of effective field theory techniques and Einstein
gravity intuition at energy scales close to the string and Planck
scales are questionable. 

New input from fundamental physics is required to address the
problems of our current models of inflation. It is likely that
superstring theory will lead to a new paradigm of early universe
cosmology. Such a new paradigm may yield a convincing realization of
inflation, but it may also give rise to a quite different scenario
for early universe cosmology.

We have presented a new cosmological background motivated by
key symmetries and making use of new degrees of freedom of
string theory. This background is characterized by an early
quasi-static Hagedorn phase during which the matter content of
the universe is a thermal gas of strings. The decay of string
winding modes leads to a smooth transition to the radiation phase
of standard cosmology, without an intervening period of inflation.
However, thermal fluctuations of strings during the Hagedorn phase
yields a new structure formation scenario. The holographic scaling
of the specific heat of a gas of closed strings on a compact
three-dimensional space then leads to an almost scale-invariant
spectrum of cosmological perturbations. A key prediction of this
scneario is a slight blue tilt of the spectrum of gravitational
waves. Note that the spectrum of the scalar modes has a slight red tilt.

There are many outstanding issues. We have presented a potential
alternative solution for the origin of structure in the universe.
Inflation, however, has other key successes. Most importantly, it
generates a large, homogeneous, high entropy and spatially flat universe starting
from initial conditions where space is small and has only a low
entropy. Are there alternatives to solving these problems which
arise from string gas cosmology? If the universe starts out cold
and large (natural initial conditions in the context of a bouncing
universe scenario), the above problems do not arise. Since string gas
cosmology may well lead to a bouncing cosmology, as in the setup of
\cite{Biswas}, this possibility should be kept in mind. Alternatively,
starting from initial conditions where all dimensions of space start
out small, there may be an initial phase of bulk dynamics which
telescopes an initial string scale into the required scale of $1 {\rm mm}$
during the Hagedorn phase (see \cite{Natalia} for an example).

Another outstanding issue is to obtain a better understanding of the
dynamics of the Hagedorn phase. Dilaton gravity does not provide
the adequate background equations, in particular at early stages during
the Hagedorn phase. An improved understanding of the background
dynamics is also required in order to be able to calculate the
temperature $T(t_i(k))$ at the time when scales $k$ exit the Hubble
radius. The value of the temperature and its k-dependence are crucial
in order to be able to calculate the magnitude of the tilt of the
power spectra of fluctuations as well as the tensor to scalar ratio. 

\centerline{\bf Acknowledgements}

I would like to thank the organizers of the Colloque 2006 of the IAP
for inviting me to speak and for their generous hospitality during this
stimulating conference. I wish to thank my collaborators Ali Nayeri, Subodh
Patil and Cumrun Vafa for a most enjoyable collaboration. I also
thank Lev Kofman, Andrei Linde and Slava Mukhanov for vigorous
discussions which helped clarify some of the points presented in
Subsection 4.5. Some of the arguments in that subsection were 
finalized after the Colloque, and appeared in \cite{Betal}. 
I am grateful to Sugumi Kanno, Jiro Soda, Damien Easson, Justin Khoury, Patrick Martineau, Ali Nayeri and Subodh Patil for their collaboration on this paper, and I  in particular thank Sugumi Kanno for allowing me
to use three figures of \cite{Betal} in this writeup. I also
with to thank Tirthabir Biswas for extensive discussions.
My research is supported by an NSERC Discovery Grant and by the Canada Research Chairs program.

%
%

%

\begin{thebibliography}{99.}
%
%
%

\bibitem{Guth}
A.~H.~Guth,
  ``The Inflationary Universe: A Possible Solution To The Horizon And Flatness
  Problems,''
  Phys.\ Rev.\ D {\bf 23}, 347 (1981).

\bibitem{Sato}
K.~Sato,
  ``First Order Phase Transition Of A Vacuum And Expansion Of The Universe,''
  Mon.\ Not.\ Roy.\ Astron.\ Soc.\  {\bf 195}, 467 (1981).

\bibitem{Starob1}
A.~A.~Starobinsky,
  ``A New Type Of Isotropic Cosmological Models Without Singularity,''
  Phys.\ Lett.\ B {\bf 91}, 99 (1980).

\bibitem{Brout}
R.~Brout, F.~Englert and E.~Gunzig,
  ``The Creation Of The Universe As A Quantum Phenomenon,''
  Annals Phys.\  {\bf 115}, 78 (1978).

\bibitem{Mukh}
V.~F.~Mukhanov and G.~V.~Chibisov,
``Quantum Fluctuation And 'Nonsingular' Universe. (In Russian),''
JETP Lett.\  {\bf 33}, 532 (1981)
[Pisma Zh.\ Eksp.\ Teor.\ Fiz.\  {\bf 33}, 549 (1981)].

\bibitem{Press}
W. Press, Phys. Scr. {\bf 21}, 702 (1980).

\bibitem{Starob2}
A.~A.~Starobinsky,
  ``Spectrum Of Relict Gravitational Radiation And The Early State Of The
  Universe,''
  JETP Lett.\  {\bf 30}, 682 (1979)
  [Pisma Zh.\ Eksp.\ Teor.\ Fiz.\  {\bf 30}, 719 (1979)].

\bibitem{Sunyaev}
R.~A.~Sunyaev and Y.~B.~Zeldovich,
  ``Small scale fluctuations of relic radiation,''
  Astrophys.\ Space Sci.\  {\bf 7}, 3 (1970).

\bibitem{Peebles}
P.~J.~E.~Peebles and J.~T.~Yu,
  ``Primeval adiabatic perturbation in an expanding universe,''
  Astrophys.\ J.\  {\bf 162}, 815 (1970).

\bibitem{COBE}
G.~F.~Smoot {\it et al.},
  ``Structure in the COBE DMR first year maps,''
  Astrophys.\ J.\  {\bf 396}, L1 (1992).

\bibitem{Boomerang}
P.~de Bernardis {\it et al.}  [Boomerang Collaboration],
  ``A Flat Universe from High-Resolution Maps of the Cosmic Microwave
  Background Radiation,''
  Nature {\bf 404}, 955 (2000)
  [arXiv:astro-ph/0004404].

\bibitem{WMAP}
C.~L.~Bennett {\it et al.},
  ``First Year Wilkinson Microwave Anisotropy Probe (WMAP) Observations:
  Preliminary Maps and Basic Results,''
  Astrophys.\ J.\ Suppl.\  {\bf 148}, 1 (2003)
  [arXiv:astro-ph/0302207].

\bibitem{Burgess}
C.~P.~Burgess,
  ``Inflatable string theory?,''
  Pramana {\bf 63}, 1269 (2004)
  [arXiv:hep-th/0408037].

\bibitem{Cline}
J.~M.~Cline,
  ``Inflation from string theory,''
  arXiv:hep-th/0501179.

\bibitem{Linde}
A.~Linde,
  ``Inflation and string cosmology,''
  eConf {\bf C040802}, L024 (2004)
  [arXiv:hep-th/0503195].

\bibitem{NBV}
A.~Nayeri, R.~H.~Brandenberger and C.~Vafa,
  ``Producing a scale-invariant spectrum of perturbations in a Hagedorn phase
  of string cosmology,''
  Phys.\ Rev.\ Lett.\  {\bf 97}, 021302 (2006)
  [arXiv:hep-th/0511140].

\bibitem{Ali}
A.~Nayeri,
   ``Inflation free, stringy generation of scale-invariant cosmological
  fluctuations in D = 3 + 1 dimensions,''
  arXiv:hep-th/0607073.

\bibitem{BNPV2}
R. H. Brandenberger, A. Nayeri, S. P. Patil and C. Vafa,
``String gas cosmology and structure formation,''
arXiv:hep-th/0608121.

\bibitem{BNPV1}
R.~H.~Brandenberger, A.~Nayeri, S.~P.~Patil and C.~Vafa,
  ``Tensor modes from a primordial Hagedorn phase of string cosmology,''
  arXiv:hep-th/0604126.

\bibitem{Betal}
R.~H.~Brandenberger {\it et al.},
  ``More on the spectrum of perturbations in string gas cosmology,''
  JCAP {\bf 0611}, 009 (2006)
  [arXiv:hep-th/0608186].

\bibitem{Sasaki}
M.~Sasaki,
  ``Large Scale Quantum Fluctuations In The Inflationary Universe,''
  Prog.\ Theor.\ Phys.\  {\bf 76}, 1036 (1986).

\bibitem{Mukh2}
V.~F.~Mukhanov,
  ``Quantum Theory Of Gauge Invariant Cosmological Perturbations,''
  Sov.\ Phys.\ JETP {\bf 67}, 1297 (1988)
  [Zh.\ Eksp.\ Teor.\ Fiz.\  {\bf 94N7}, 1 (1988)].

\bibitem{MFB}
V.~F.~Mukhanov, H.~A.~Feldman and R.~H.~Brandenberger,
``Theory of cosmological perturbations. Part 1. Classical perturbations. Part
  2. Quantum theory of perturbations. Part 3. Extensions,''
Phys.~Rept.~{\bf 215}, 203 (1992).

\bibitem{RHBrev2}
R.~H.~Brandenberger,
  ``Lectures on the theory of cosmological perturbations,''
  Lect.\ Notes Phys.\  {\bf 646}, 127 (2004)
  [arXiv:hep-th/0306071].

\bibitem{acoustic}
V.~Mukhanov,
  ``CMB-slow, or How to Estimate Cosmological Parameters by Hand,''
  arXiv:astro-ph/0303072.

\bibitem{chaotic}
A.~D.~Linde,
  ``Chaotic Inflation,''
  Phys.\ Lett.\ B {\bf 129}, 177 (1983).

\bibitem{hybrid}
A.~D.~Linde,
  ``Hybrid inflation,''
  Phys.\ Rev.\ D {\bf 49}, 748 (1994)
  [arXiv:astro-ph/9307002].

\bibitem{kung}
R.~H.~Brandenberger and J.~H.~Kung,
  ``Chaotic Inflation As An Attractor In Initial Condition Space,''
  Phys.\ Rev.\ D {\bf 42}, 1008 (1990).

\bibitem{hume}
H.~A.~Feldman and R.~H.~Brandenberger,
  ``Chaotic Inflation With Metric And Matter Perturbations,''
  Phys.\ Lett.\ B {\bf 227}, 359 (1989).

\bibitem{goldwirth}
D.~S.~Goldwirth and T.~Piran,
  ``Initial conditions for inflation,''
  Phys.\ Rept.\  {\bf 214}, 223 (1992).

\bibitem{new}
A.~D.~Linde,
  ``A New Inflationary Universe Scenario: A Possible Solution Of The Horizon,
  Flatness, Homogeneity, Isotropy And Primordial Monopole Problems,''
  Phys.\ Lett.\ B {\bf 108}, 389 (1982);\\
A.~Albrecht and P.~J.~Steinhardt,
  ``Cosmology For Grand Unified Theories With Radiatively Induced Symmetry
  Breaking,''
  Phys.\ Rev.\ Lett.\  {\bf 48}, 1220 (1982).

\bibitem{ghazalscott}
R.~Brandenberger, G.~Geshnizjani and S.~Watson,  
  ``On the initial conditions for brane inflation,''  
  Phys.\ Rev.\ D {\bf 67}, 123510 (2003)
  [arXiv:hep-th/0302222].  

\bibitem{eternal}
A.~D.~Linde,
  ``Eternal Chaotic Inflation,''  
  Mod.\ Phys.\ Lett.\ A {\bf 1}, 81 (1986);\\
A.~D.~Linde and D.~A.~Linde,
  ``Topological defects as seeds for eternal inflation,''
  Phys.\ Rev.\ D {\bf 50}, 2456 (1994)
  [arXiv:hep-th/9402115].

\bibitem{MSSM}
R.~Allahverdi, K.~Enqvist, J.~Garcia-Bellido and A.~Mazumdar,
  ``Gauge invariant MSSM inflaton,''
  arXiv:hep-ph/0605035.

\bibitem{Adams}
F.~C.~Adams, K.~Freese and A.~H.~Guth,
  ``Constraints On The Scalar Field Potential In Inflationary Models,''
  Phys.\ Rev.\ D {\bf 43}, 965 (1991).

\bibitem{VilShell}
A. Vilenkin and E.P.S. Shellard, \textit{Cosmic Strings and other
Topological Defects} (Cambridge Univ. Press, Cambridge, 1994).

\bibitem{HK}
M.~B.~Hindmarsh and T.~W.~B.~Kibble,
  ``Cosmic strings,''
  Rept.\ Prog.\ Phys.\  {\bf 58}, 477 (1995)
  [arXiv:hep-ph/9411342].

\bibitem{RHBrev0}
R.~H.~Brandenberger,
  ``Topological defects and structure formation,''
  Int.\ J.\ Mod.\ Phys.\ A {\bf 9}, 2117 (1994)
  [arXiv:astro-ph/9310041].

\bibitem{CSpapers}
N.~Turok and R.~H.~Brandenberger,
  ``Cosmic Strings And The Formation Of Galaxies And Clusters Of Galaxies,''
  Phys.\ Rev.\ D {\bf 33}, 2175 (1986);\\
H. Sato, ``Galaxy Formation by Cosmic Strings,''
  Prog. Theor. Phys.\  {\bf 75}, 1342 (1986);\\
A. Stebbins, ``Cosmic Strings and Cold Matter'',
  Ap. J. (Lett.) {\bf 303}, L21 (1986).

\bibitem{CSanis}
A.~Albrecht, D.~Coulson, P.~Ferreira and J.~Magueijo,
  ``Causality and the microwave background,''
  Phys.\ Rev.\ Lett.\  {\bf 76}, 1413 (1996)
  [arXiv:astro-ph/9505030];\\
J.~Magueijo, A.~Albrecht, D.~Coulson and P.~Ferreira,
  ``Doppler peaks from active perturbations,''
  Phys.\ Rev.\ Lett.\  {\bf 76}, 2617 (1996)
  [arXiv:astro-ph/9511042];\\
U.~L.~Pen, U.~Seljak and N.~Turok,
  ``Power spectra in global defect theories of cosmic structure formation,''
  Phys.\ Rev.\ Lett.\  {\bf 79}, 1611 (1997)
  [arXiv:astro-ph/9704165].

\bibitem{RHBrev1}
R.~H.~Brandenberger,  
``Inflationary cosmology: Progress and problems,'' publ. in proc. of
IPM School On Cosmology 1999: Large Scale Structure Formation,
  arXiv:hep-ph/9910410.  

\bibitem{Jerome1}
R.~H.~Brandenberger and J.~Martin, ``The robustness of inflation to changes in super-Planck-scale physics,''
Mod.~Phys.~Lett.~A~{\bf 16}, 999 (2001), 
[arXiv:astro-ph/0005432];\\
J.~Martin and R.~H.~Brandenberger,
``The trans-Planckian problem of inflationary cosmology,''
Phys.~Rev.~D~{\bf 63}, 123501 (2001), 
[arXiv:hep-th/0005209].

\bibitem{Niemeyer}
J.~C.~Niemeyer,
``Inflation with a high frequency cutoff,''
Phys.~Rev.~D~{\bf 63}, 123502 (2001),
[arXiv:astro-ph/0005533]; \\
S.~Shankaranarayanan, 
``Is there an imprint of Planck scale physics on inflationary cosmology?,''
Class.~Quant.~Grav.~{\bf 20}, 75 (2003), 
[arXiv:gr-qc/0203060];\\
J.~C.~Niemeyer and R.~Parentani,
  ``Trans-Planckian dispersion and scale-invariance of inflationary
  perturbations,''
 Phys.~Rev.~D~{\bf 64}, 101301 (2001),
[arXiv:astro-ph/0101451].

\bibitem{Unruh}
W.~G.~Unruh,
  ``Sonic analog of black holes and the effects of high frequencies on black
  hole evaporation,''
  Phys.\ Rev.\ D {\bf 51}, 2827 (1995).

\bibitem{CJ}
S.~Corley and T.~Jacobson,
  ``Hawking Spectrum and High Frequency Dispersion,''
  Phys.\ Rev.\ D {\bf 54}, 1568 (1996)
  [arXiv:hep-th/9601073].

\bibitem{Jerome3}
R.~H.~Brandenberger and J.~Martin,
  ``Back-reaction and the trans-Planckian problem of inflation revisited,''
  Phys.\ Rev.\ D {\bf 71}, 023504 (2005)
  [arXiv:hep-th/0410223].

\bibitem{Tanaka}
T.~Tanaka, 
``A comment on trans-Planckian physics in inflationary universe,''
[arXiv:astro-ph/0012431].

\bibitem{Starob3}
A.~A.~Starobinsky, 
``Robustness of the inflationary perturbation spectrum to trans-Planckian   
physics,''
Pisma Zh.~Eksp.~Teor.~Fiz.~{\bf 73}, 415 (2001),
[JETP Lett.\ {\bf 73}, 371 (2001)], [arXiv:astro-ph/0104043].

\bibitem{Easther}
C.~S.~Chu, B.~R.~Greene and G.~Shiu, 
``Remarks on inflation and noncommutative geometry,''
Mod.~Phys.~Lett.~A~{\bf 16}, 2231
(2001), [arXiv:hep-th/0011241]; \\
R.~Easther, B.~R.~Greene, W.~H.~Kinney and G.~Shiu, 
``Inflation as a probe of short distance physics,''
Phys.~Rev.~D~{\bf 64}, 103502 (2001),
[arXiv:hep-th/0104102]; \\
R.~Easther, B.~R.~Greene, W.~H.~Kinney and G.~Shiu, 
``Imprints of short distance physics on inflationary cosmology,''
Phys.~Rev.~D~{\bf 67}, 063508 (2003), [arXiv:hep-th/0110226];\\
F.~Lizzi, G.~Mangano, G.~Miele and M.~Peloso, 
``Cosmological perturbations and short distance physics from  noncommutative   
geometry,''
JHEP~{\bf 0206}, 049 (2002) 
[arXiv:hep-th/0203099];\\
S.~F.~Hassan and M.~S.~Sloth, 
``Trans-Planckian effects in inflationary cosmology and the modified   
uncertainty principle,''
Nucl.~Phys.~B~{\bf 674}, 434 (2003),
[arXiv:hep-th/0204110].

\bibitem{Ho}
R.~Brandenberger and P.~M.~Ho,
``Noncommutative spacetime, stringy spacetime uncertainty principle, and   
density fluctuations,''
 Phys.~Rev.~D~{\bf 66}, 023517 (2002),
[AAPPS Bull.~{\bf 12N1}, 10 (2002)], [arXiv:hep-th/0203119].

\bibitem{Kempf}
A.~Kempf and J.~C.~Niemeyer, 
``Perturbation spectrum in inflation with cutoff,''
Phys.~Rev.~D~{\bf 64}, 103501 (2001),
[arXiv:astro-ph/0103225].

\bibitem{minimal}
U.~H.~Danielsson, Phys.~Rev.~D~{\bf 66}, 023511 (2002),
``A note on inflation and transplanckian physics,''
[arXiv:hep-th/0203198];\\
V.~Bozza, M.~Giovannini and G.~Veneziano, JCAP~{\bf 0305}, 001 (2003),
``Cosmological perturbations from a new-physics hypersurface,''
[arXiv:hep-th/0302184];\\
J.~C.~Niemeyer, R.~Parentani and D.~Campo,
  ``Minimal modifications of the primordial power spectrum from an  adiabatic
  short distance cutoff,''
  Phys.\ Rev.\ D {\bf 66}, 083510 (2002)
  [arXiv:hep-th/0206149].

\bibitem{Cliff}
C.~P.~Burgess, J.~M.~Cline, F.~Lemieux and R.~Holman,
  ``Are inflationary predictions sensitive to very high energy physics?,''
  JHEP {\bf 0302}, 048 (2003)
  [arXiv:hep-th/0210233];\\
K.~Schalm, G.~Shiu and J.~P.~van der Schaar,
  ``The cosmological vacuum ambiguity, effective actions, and  transplanckian
  effects in inflation,''
  AIP Conf.\ Proc.\  {\bf 743}, 362 (2005)
  [arXiv:hep-th/0412288].

\bibitem{Jerome2}
J.~Martin and R.~Brandenberger,
  ``On the dependence of the spectra of fluctuations in inflationary  cosmology
  on trans-Planckian physics,''
  Phys.\ Rev.\ D  {\bf 68}, 063513 (2003)
  [arXiv:hep-th/0305161].

\bibitem{HE}
S. Hawking and G. Ellis, \textit{The Large-Scale Structure of Space-Time}
(Cambridge Univ. Press, Cambridge, 1973).

\bibitem{Borde}
A.~Borde and A.~Vilenkin,
  ``Eternal inflation and the initial singularity,''
  Phys.\ Rev.\ Lett.\  {\bf 72}, 3305 (1994)
  [arXiv:gr-qc/9312022].

\bibitem{swamp} N.~Arkani-Hamed, L.~Motl, A.~Nicolis and C.~Vafa,
  ``The string landscape, black holes and gravity as the weakest force,''
  arXiv:hep-th/0601001.

\bibitem{BMS}
R.~H.~Brandenberger, V.~F.~Mukhanov and A.~Sornborger,
  ``A Cosmological theory without singularities,''
  Phys.\ Rev.\ D {\bf 48}, 1629 (1993)
  [arXiv:gr-qc/9303001];\\
V.~F.~Mukhanov and R.~H.~Brandenberger,
  ``A Nonsingular universe,''
  Phys.\ Rev.\ Lett.\  {\bf 68}, 1969 (1992).

\bibitem{Biswas}
T.~Biswas, A.~Mazumdar and W.~Siegel,
  ``Bouncing universes in string-inspired gravity,''
  JCAP {\bf 0603}, 009 (2006)
  [arXiv:hep-th/0508194].

\bibitem{Bojowald}
M.~Bojowald,   
``Loop quantum cosmology,''   
Living Rev.\ Rel.\  {\bf 8}, 11 (2005)   
[arXiv:gr-qc/0601085].   

\bibitem{Hagedorn}
R.~Hagedorn,
  ``Statistical Thermodynamics Of Strong Interactions At High-Energies,''
  Nuovo Cim.\ Suppl.\  {\bf 3}, 147 (1965).

\bibitem{BV}
R.~H.~Brandenberger and C.~Vafa,
  ``Superstrings In The Early Universe,''
  Nucl.\ Phys.\ B {\bf 316}, 391 (1989).

\bibitem{RHBrev4}
R.~H.~Brandenberger,
``Back reaction of cosmological perturbations and the cosmological constant
problem,'' publ. in the proc. of the
18th IAP Colloquium On The Nature Of Dark Energy: Observational And Theoretical Results On The Accelerating Universe,
arXiv:hep-th/0210165.

\bibitem{branereview}
P.~Brax, C.~van de Bruck and A.~C.~Davis,
  ``Brane world cosmology,''
  Rept.\ Prog.\ Phys.\  {\bf 67}, 2183 (2004)
  [arXiv:hep-th/0404011].

\bibitem{Mahbub}
M.~Majumdar and A.-C. Davis,
  ``Cosmological creation of D-branes and anti-D-branes,''
  JHEP {\bf 0203}, 056 (2002)
  [arXiv:hep-th/0202148].

\bibitem{Mairi}
R.~Durrer, M.~Kunz and M.~Sakellariadou,
  ``Why do we live in 3+1 dimensions?,''
  Phys.\ Lett.\ B {\bf 614}, 125 (2005)
  [arXiv:hep-th/0501163].

\bibitem{Lisa}
A.~Karch and L.~Randall,
  ``Relaxing to three dimensions,''
  Phys.\ Rev.\ Lett.\  {\bf 95}, 161601 (2005)
  [arXiv:hep-th/0506053].

\bibitem{TV}
A.~A.~Tseytlin and C.~Vafa,
  ``Elements of string cosmology,''
  Nucl.\ Phys.\ B {\bf 372}, 443 (1992)
  [arXiv:hep-th/9109048].

\bibitem{ABE}
S.~Alexander, R.~H.~Brandenberger and D.~Easson,
  ``Brane gases in the early universe,''
  Phys.\ Rev.\ D {\bf 62}, 103509 (2000)
  [arXiv:hep-th/0005212].

\bibitem{Perlt}
J.~Kripfganz and H.~Perlt,
  ``Cosmological Impact Of Winding Strings,''
  Class.\ Quant.\ Grav.\  {\bf 5}, 453 (1988).

\bibitem{BattWat}
T.~Battefeld and S.~Watson,   
``String gas cosmology,''   
Rev.\ Mod.\ Phys.\  {\bf 78}, 435 (2006)   
[arXiv:hep-th/0510022].   

\bibitem{RHBrev5}
R.~H.~Brandenberger,   
``Challenges for string gas cosmology,'' publ. in proc. of the   
59th Yamada Conference On Inflating Horizon Of Particle Astrophysics And Cosmology, 
arXiv:hep-th/0509099.   

\bibitem{RHBrev6}
R.~H.~Brandenberger,   
``Moduli stabilization in string gas cosmology,''   
Prog.\ Theor.\ Phys.\ Suppl.\  {\bf 163}, 358 (2006)   
[arXiv:hep-th/0509159].   

\bibitem{Pol}
J. Polchinski, \textit{String Theory, Vols. 1 and 2},
(Cambridge Univ. Press, Cambridge, 1998).

\bibitem{Boehm}
T.~Boehm and R.~Brandenberger,
  ``On T-duality in brane gas cosmology,''
  JCAP {\bf 0306}, 008 (2003)
  [arXiv:hep-th/0208188].

\bibitem{Biswas2}
T.~Biswas, R.~Brandenberger, A.~Mazumdar and W.~Siegel,   
``Non-perturbative gravity, Hagedorn bounce and CMB,''   
arXiv:hep-th/0610274.   

\bibitem{gellis}
G.~F.~R.~Ellis and R.~Maartens,   
  ``The emergent universe: Inflationary cosmology with no singularity,''      
  Class.\ Quant.\ Grav.\  {\bf 21}, 223 (2004)   
  [arXiv:gr-qc/0211082].   

\bibitem{KKLM}
N.~Kaloper, L.~Kofman, A.~Linde and V.~Mukhanov,   
``On the new string theory inspired mechanism of generation of cosmological   
perturbations,''   
JCAP {\bf 0610}, 006 (2006)   
[arXiv:hep-th/0608200].   

\bibitem{Kabat3}
R.~Easther, B.~R.~Greene, M.~G.~Jackson and D.~Kabat,
  ``String windings in the early universe,''
  JCAP {\bf 0502}, 009 (2005)
  [arXiv:hep-th/0409121].

\bibitem{Danos}
R.~Danos, A.~R.~Frey and A.~Mazumdar,
  ``Interaction rates in string gas cosmology,''
  Phys.\ Rev.\ D {\bf 70}, 106010 (2004)
  [arXiv:hep-th/0409162].

\bibitem{Watson1}
S.~Watson and R.~H.~Brandenberger,
  ``Isotropization in brane gas cosmology,''
  Phys.\ Rev.\ D {\bf 67}, 043510 (2003)
  [arXiv:hep-th/0207168].

\bibitem{Ven}
G.~Veneziano,
  ``Scale factor duality for classical and quantum strings,''
  Phys.\ Lett.\ B {\bf 265}, 287 (1991).

\bibitem{Riotto}
M.~Maggiore and A.~Riotto,
  ``D-branes and cosmology,''
  Nucl.\ Phys.\ B {\bf 548}, 427 (1999)
  [arXiv:hep-th/9811089].

\bibitem{Kaya06}
S.~P.~Patil,
  ``Moduli (dilaton, volume and shape) stabilization via massless F and D
  arXiv:hep-th/0504145;\\
S.~Cremonini and S.~Watson,
  ``Dilaton dynamics from production of tensionless membranes,''
  Phys.\ Rev.\ D {\bf 73}, 086007 (2006)
  [arXiv:hep-th/0601082];\\
S.~Arapoglu, A.~Karakci and A.~Kaya,
  ``S-duality in string gas cosmology,''
  arXiv:hep-th/0611193.

\bibitem{Watson2}
S.~Watson and R.~Brandenberger,
  ``Stabilization of extra dimensions at tree level,''
  JCAP {\bf 0311}, 008 (2003)
  [arXiv:hep-th/0307044].

\bibitem{Subodh1}
S.~P.~Patil and R.~Brandenberger,
  ``Radion stabilization by stringy effects in general relativity and  dilaton
  gravity,''
  Phys.\ Rev.\ D {\bf 71}, 103522 (2005)
  [arXiv:hep-th/0401037].

\bibitem{Subodh2}
S.~P.~Patil and R.~H.~Brandenberger,
  ``The cosmology of massless string modes,''
  JCAP {\bf 0601}, 005 (2006)
  [arXiv:hep-th/0502069].

\bibitem{BattWat2}
T.~Battefeld and S.~Watson,
  ``Effective field theory approach to string gas cosmology,''
  JCAP {\bf 0406}, 001 (2004)
  [arXiv:hep-th/0403075].

\bibitem{Edna}
R.~Brandenberger, Y.~K.~Cheung and S.~Watson,
  ``Moduli stabilization with string gases and fluxes,''
  JHEP {\bf 0605}, 025 (2006)
  [arXiv:hep-th/0501032].

\bibitem{Sugumi}
S. Kanno and J. Soda,
``Moduli Stabilization in String Gas Cosmology,''
Phys.\ Rev.\ D {\bf 72}, 104023 (2005)
[arXiv:hep-th/0509074].

\bibitem{Watson3}
S.~Watson,
  ``Moduli stabilization with the string Higgs effect,''
  Phys.\ Rev.\ D {\bf 70}, 066005 (2004)
  [arXiv:hep-th/0404177].

\bibitem{Eva}
L.~Kofman, A.~Linde, X.~Liu, A.~Maloney, L.~McAllister and E.~Silverstein,
  ``Beauty is attractive: Moduli trapping at enhanced symmetry points,''
  JHEP {\bf 0405}, 030 (2004)
  [arXiv:hep-th/0403001].

\bibitem{other}
G.~B.~Cleaver and P.~J.~Rosenthal, 
  ``String cosmology and the dimension of space-time,'' 
  Nucl.\ Phys.\ B {\bf 457}, 621 (1995) 
  [arXiv:hep-th/9402088];\\ 
M.~Sakellariadou, 
 ``Numerical Experiments in String Cosmology,'' 
 Nucl.\ Phys.\ B {\bf 468}, 319 (1996) 
 [arXiv:hep-th/9511075];\\ 
R.~Easther, B.~R.~Greene and M.~G.~Jackson,
  ``Cosmological string gas on orbifolds,''
  Phys.\ Rev.\ D {\bf 66}, 023502 (2002)
  [arXiv:hep-th/0204099]l\\
R.~Easther, B.~R.~Greene, M.~G.~Jackson and D.~Kabat,
``Brane gas cosmology in M-theory: Late time behavior,''
Phys.\ Rev.\ D {\bf 67}, 123501 (2003)
[arXiv:hep-th/0211124];\\
R.~Easther, B.~R.~Greene, M.~G.~Jackson and D.~Kabat,
``Brane gases in the early universe: Thermodynamics and cosmology,''
JCAP {\bf 0401}, 006 (2004)
[arXiv:hep-th/0307233];\\
S.~H.~S.~Alexander,
  ``Brane gas cosmology, M-theory and little string theory,''
  JHEP {\bf 0310}, 013 (2003)
  [arXiv:hep-th/0212151];\\
B.~A.~Bassett, M.~Borunda, M.~Serone and S.~Tsujikawa,
  ``Aspects of string-gas cosmology at finite temperature,''
  Phys.\ Rev.\ D {\bf 67}, 123506 (2003)
  [arXiv:hep-th/0301180];\\
D.~A.~Easson,
``Brane gases on K3 and Calabi-Yau manifolds,''
Int.\ J.\ Mod.\ Phys.\ A {\bf 18}, 4295 (2003)
[arXiv:hep-th/0110225];\\
A.~Campos,
  ``Late-time dynamics of brane gas cosmology,''
  Phys.\ Rev.\ D {\bf 68}, 104017 (2003)
  [arXiv:hep-th/0304216];\\
T.~Biswas,
``Cosmology with branes wrapping curved internal manifolds,''
JHEP {\bf 0402}, 039 (2004)
[arXiv:hep-th/0311076];\\
A.~Kaya and T.~Rador,
``Wrapped branes and compact extra dimensions in cosmology,''
Phys.\ Lett.\ B {\bf 565}, 19 (2003)
[arXiv:hep-th/0301031];\\
A.~Kaya,
``On winding branes and cosmological evolution of extra dimensions in  string
theory,''
Class.\ Quant.\ Grav.\  {\bf 20}, 4533 (2003)
[arXiv:hep-th/0302118];\\
A.~Campos,
``Late cosmology of brane gases with a two-form field,''
Phys.\ Lett.\ B {\bf 586}, 133 (2004)
[arXiv:hep-th/0311144];\\
S.~Watson and R.~Brandenberger,
``Linear perturbations in brane gas cosmology,''
JHEP {\bf 0403}, 045 (2004)
[arXiv:hep-th/0312097];\\
S.~Watson,
``UV perturbations in brane gas cosmology,''
Phys.\ Rev.\ D {\bf 70}, 023516 (2004)
[arXiv:hep-th/0402015];\\
R.~Brandenberger, D.~A.~Easson and A.~Mazumdar, 
 ``Inflation and brane gases,'' 
 Phys.\ Rev.\ D {\bf 71}, 083514 (2005)
 [arXiv:hep-th/0307043];\\ 
A.~Kaya,
``Volume stabilization and acceleration in brane gas cosmology,''
JCAP {\bf 0408}, 014 (2004)
[arXiv:hep-th/0405099];\\
S.~Arapoglu and A.~Kaya,
``D-brane gases and stabilization of extra dimensions in dilaton gravity,''
Phys.\ Lett.\ B {\bf 603}, 107 (2004)
[arXiv:hep-th/0409094];\\
T.~Rador, 
 ``Intersection democracy for winding branes and stabilization of extra 
 dimensions,'' 
 Phys.\ Lett.\ B {\bf 621}, 176 (2005) 
 [arXiv:hep-th/0501249];\\ 
T. Rador, 
``Vibrating winding branes, wrapping democracy and stabilization of extra 
 dimensions in dilaton gravity,'' 
 JHEP {\bf 0506}, 001 (2005) 
 [arXiv:hep-th/0502039];\\ 
T. Rador, 
``Stabilization of extra dimensions and the dimensionality of the  observed 
 space,'' 
 arXiv:hep-th/0504047;\\ 
A.~Kaya,  
``Brane gases and stabilization of shape moduli with momentum and winding 
 stress,''  
Phys.\ Rev.\ D {\bf 72}, 066006 (2005) 
 [arXiv:hep-th/0504208];\\ 
F.~Ferrer and S.~Rasanen,
 ``Dark energy and decompactification in string gas cosmology,''   
JHEP {\bf 0602}, 016 (2006)   
[arXiv:hep-th/0509225];\\   
M.~Borunda and L.~Boubekeur, 
 ``The effect of alpha' corrections in string gas cosmology,'' 
 JCAP {\bf 0610}, 002 (2006)
 [arXiv:hep-th/0604085];\\  
A.~Chatrabhuti, 
 ``Target space duality and moduli stabilization in string gas cosmology,'' 
 arXiv:hep-th/0602031;\\
J.~Y.~Kim,   
``Stabilization of the extra dimensions in brane gas cosmology with bulk   
flux,''   
arXiv:hep-th/0608131.  

\bibitem{dilaton}
A.~J.~Berndsen and J.~M.~Cline,
  ``Dilaton stabilization in brane gas cosmology,''
  Int.\ J.\ Mod.\ Phys.\ A {\bf 19}, 5311 (2004)
  [arXiv:hep-th/0408185];\\
A.~Berndsen, T.~Biswas and J.~M.~Cline,
  ``Moduli stabilization in brane gas cosmology with superpotentials,''
  JCAP {\bf 0508}, 012 (2005)
  [arXiv:hep-th/0505151];\\
D.~A.~Easson and M.~Trodden,
  ``Moduli stabilization and inflation using wrapped branes,''
  Phys.\ Rev.\ D {\bf 72}, 026002 (2005)
  [arXiv:hep-th/0505098];\\
S.~Cremonini and S.~Watson, 
 ``Dilaton dynamics from production of tensionless membranes,''
  Phys.\ Rev.\ D {\bf 73}, 086007 (2006)   
[arXiv:hep-th/0601082].

\bibitem{Subodh3}
S.~P.~Patil,
  ``Moduli (dilaton, volume and shape) stabilization via massless F and D
  string modes,''
  arXiv:hep-th/0504145.

\bibitem{BEK}
R.~Brandenberger, D.~A.~Easson and D.~Kimberly,
  ``Loitering phase in brane gas cosmology,''
  Nucl.\ Phys.\ B {\bf 623}, 421 (2002)
  [arXiv:hep-th/0109165].

\bibitem{Natalia}
R.~Brandenberger and N.~Shuhmaher,
  ``The Confining Heterotic Brane Gas: A Non-Inflationary Solution to the
  Entropy and Horizon Problems of Standard Cosmology,''
  JHEP {\bf 0601}, 074 (2006)
  [arXiv:hep-th/0511299].

\bibitem{Deo}
 N.~Deo, S.~Jain, O.~Narayan and C.~I.~Tan,
  ``The Effect of topology on the thermodynamic limit for a string gas,''
  Phys.\ Rev.\ D {\bf 45}, 3641 (1992).

\bibitem{BST}
J.~M.~Bardeen, P.~J.~Steinhardt and M.~S.~Turner,
 ``Spontaneous Creation Of Almost Scale - Free Density Perturbations In An
 Inflationary Universe,''
 Phys.\ Rev.\ D {\bf 28}, 679 (1983).

\bibitem{BK}
R.~H.~Brandenberger and R.~Kahn,
 ``Cosmological Perturbations In Inflationary Universe Models,''
 Phys.\ Rev.\ D {\bf 29}, 2172 (1984);\\
R.~H.~Brandenberger,
 ``Quantum Fluctuations As The Source Of Classical Gravitational Perturbations
 In The Inflationary Universe,''
 Nucl.\ Phys.\ B {\bf 245}, 328 (1984).

\end{thebibliography}
%



\printindex
\end{document}